\documentclass[10pt]{article}
\usepackage{multicol}
\usepackage{graphicx}
\usepackage{amsmath}
\usepackage{CJK}

\begin{document}

\begin{center}

{\Large \bf Kinetic freeze-out temperatures in central and
peripheral collisions: Which one is larger?}

\vskip.5cm

\begin{CJK*}{GBK}{song}
Hai-Ling Lao, Fu-Hu Liu{\footnote{E-mail: fuhuliu@163.com;
fuhuliu@sxu.edu.cn}}, Bao-Chun Li, Mai-Ying Duan
\end{CJK*}

{\small\it Institute of Theoretical Physics \& State Key
Laboratory of Quantum Optics and Quantum Optics Devices,

Shanxi University, Taiyuan, Shanxi 030006, China}
\end{center}

\vskip.25cm

{\bf Abstract:} The kinetic freeze-out temperatures, $T_0$, in
nucleus-nucleus collisions at the Relativistic Heavy Ion Collider
(RHIC) and Large Hadron Collider (LHC) energies are extracted by
four methods: i) the Blast-Wave model with Boltzmann-Gibbs
statistics (the BGBW model), ii) the Blast-Wave model with Tsallis
statistics (the TBW model), iii) the Tsallis distribution with
flow effect (the improved Tsallis distribution), and iv) the
intercept in $T=T_0+am_0$ (the alternative method), where $m_0$
denotes the rest mass and $T$ denotes the effective temperature
which can be obtained by different distribution functions. It is
found that the relative sizes of $T_0$ in central and peripheral
collisions obtained by the conventional BGBW model which uses a
zero or nearly zero transverse flow velocity, $\beta_T$, are
contradictory in tendency with other methods. With a
re-examination for $\beta_T$ in the first method in which
$\beta_T$ is taken to be $\sim(0.40\pm0.07)c$, a recalculation
presents a consistent result with others. Finally, our results
show that the kinetic freeze-out temperature in central collisions
is larger than that in peripheral collisions.
\\

\\


\vskip1.0cm

\begin{multicols}{2}

{\section{Introduction}}

Temperature is an important concept in high energy nucleus-nucleus
collisions. Usually, three types of temperatures which contain the
chemical freeze-out temperature, kinetic freeze-out temperature,
and effective temperature are used in literature [1--5]. The
chemical freeze-out temperature describes the excitation degree of
the interacting system at the stage of chemical equilibrium in
which the chemical components (relative fractions) of particles
are fixed. The kinetic freeze-out temperature describes the
excitation degree of the interacting system at the stage of
kinetic and thermal equilibrium in which the (transverse) momentum
spectra of particles are no longer changed. The effective
temperature is not a real temperature. In fact, the effective
temperature is related to particle mass and can be extracted from
the transverse momentum spectra by using some distribution laws
such as the standard (Boltzmann, Fermi-Dirac, and Bose-Einstein),
Tsallis, and so forth.

Generally, the chemical freeze-out temperature is usually obtained
from the particle ratios [6--8]. It is equal to or larger than the
kinetic freeze-out temperature due to the the chemical equilibrium
being meanwhile or earlier than the kinetic equilibrium. The
effective temperature is larger than the kinetic freeze-out
temperature due to mass and flow effects [9, 10]. Both the
chemical freeze-out and effective temperatures in central
nucleus-nucleus collisions are larger than those in peripheral
collisions due to more violent interactions occurring in central
collisions. In fact, central collisions contain more nucleons, and
peripheral collisions contains less nucleons. Usually, there are
small dissents in the extractions of chemical freeze-out
temperature and effective temperature. As for the extraction of
kinetic freeze-out temperature, the situations are largely
non-uniform.

Currently, four main methods are used in the extraction of kinetic
freeze-out temperature $T_0$, which are i) the Blast-Wave model
with Boltzmann-Gibbs statistics (the BGBW model) [11--13], ii) the
Blast-Wave model with Tsallis statistics (the TBW model) [14],
iii) the Tsallis distribution with flow effect (the improved
Tsallis distribution) [15, 16], and iv) the intercept in
$T=T_0+am_0$ (the alternative method) [12, 17--20], where $m_0$
denotes the rest mass and $T$ denotes the effective temperature
which can be obtained by different distribution functions. In
detail, the alternative method can be divided into a few
sub-methods due to different distributions being used. Generally,
we are inclined to use the standard and Tsallis distributions in
the alternative method due to the standard distribution being
closest to the ideal gas model in thermodynamics, and the Tsallis
distribution describing a wide spectrum which needs two- or
three-component standard distribution to be fitted [21].

The kinetic freeze-out temperature $T_0$ and the mean transverse
radial flow velocity $\beta_T$ can be simultaneously extracted by
the first three methods. The alternative method needs further
treatments in extracting the flow velocity. In our recent works
[22--24], the mean transverse flow velocity $\beta_T$ is regarded
as the slope in the relation $\langle p_T \rangle=\langle p_T
\rangle_0+ \beta_T \overline{m}$, where $\langle p_T \rangle$
denotes the mean value of transverse momenta $p_T$, $\langle p_T
\rangle_0$ denotes the mean transverse momentum in the case of
zero flow velocity, and $\overline{m}$ denotes the mean moving
mass. The mean flow velocity $\beta$ is regarded as the slope in
the relation $\langle p \rangle=\langle p \rangle_0+ \beta
\overline{m}$, where $\langle p \rangle$ denotes the mean value of
momenta $p$ and $\langle p \rangle_0$ denotes the mean momentum in
the case of zero flow velocity. Although the mean transverse
radial flow and mean transverse flow are not exactly the same, we
use the same symbol to denote their velocities and neglect the
difference between them. In fact, the mean transverse radial flow
contains only the isotropic flow, and the mean transverse flow
contains both the isotropic and anisotropic flows. The isotropic
flow is mainly caused by isotropic expansion of the interacting
system, and the anisotropic flow is mainly caused by anisotropic
squeeze between two incoming nuclei.

We are interested in the coincidence and difference among the four
methods in the extractions of $T_0$ and $\beta_T$. In this paper,
we shall use the four methods to extract $T_0$ and $\beta_T$ from
the $p_T$ spectra of identified particles produced in central and
peripheral gold-gold (Au-Au) collisions at the center-of-mass
energy per nucleon pair $\sqrt{s_{NN}}=200$ GeV (the top RHIC
energy) and in central and peripheral lead-lead (Pb-Pb) collisions
at $\sqrt{s_{NN}}=2.76$ TeV (one of the LHC energies). The model
results on the $p_T$ spectra are compared with the experimental
data of the PHENIX [25], STAR [26, 27], and ALICE Collaborations
[28, 29], and the model results on $T_0$ and $\beta_T$ in
different collisions and by different methods are compared each
other.

The rest part of this paper is structured as follows. The
formalism and method are shortly described in section 2. Results
and discussion are given in section 3. Finally, we summarize our
main observations and conclusions in section 4.
\\

{\section{Formalism and method}}

The four methods can be found in related references [11--20]. To
give a whole representation of this paper, we present directly and
concisely the four methods in the following. In the
representation, some quantities such as the kinetic freeze-out
temperature, the mean transverse (radial) flow velocity, and the
effective temperature in different methods are uniformly denoted
by $T_0$, $\beta_T$, and $T$, respectively, though different
methods correspond to different values. All of the model
descriptions are presented at the mid-rapidity which uses the
rapidity $y \approx 0$ and results in $\cosh(y) \approx 1$ which
appears in some methods. At the same time, the spin property and
chemical potential in the $p_T$ spectra are neglected due to their
small influences in high energy collisions. This means that we can
give up the Fermi-Dirac and Bose-Einstein distributions, and use
only the Boltzmann distribution in the case of considering the
standard distribution.

According to refs. [11--13], the BGBW model results in the $p_T$
distribution to be
\begin{align}
f_1(p_T)=&C_1 p_T m_T \int_0^R rdr \times \nonumber\\
& I_0 \bigg[\frac{p_T \sinh(\rho)}{T_0} \bigg] K_1 \bigg[\frac{m_T
\cosh(\rho)}{T_0} \bigg],
\end{align}
where $C_1$ is the normalized constant which results in
$\int_0^{\infty} f_1(p_T)dp_T=1$, $I_0$ and $K_1$ are the modified
Bessel functions of the first and second kinds respectively,
$m_T=\sqrt{p_T^2+m_0^2}$ is the transverse mass, $\rho= \tanh^{-1}
[\beta(r)]$ is the boost angle, $\beta(r)= \beta_S(r/R)^{n_0}$ is
a self-similar flow profile, $\beta_S$ is the flow velocity on the
surface of the thermal source, $r/R$ is the relative radial
position in the thermal source, and $n_0$ is a free parameter
which is customarily chosen to be 2 [11] due to the quadratic
profile resembling the solutions of hydrodynamics closest [30].
Generally, $\beta_T=(2/R^2)\int_0^R r\beta(r)dr
=2\beta_S/(n_0+2)$. In the case of $n_0=2$ as used in ref. [11],
we have $\beta_T=0.5\beta_S$ [31].

According to refs. [14], the TBW model results in the $p_T$
distribution to be
\begin{align}
f_2(p_T)&=C_2 p_T m_T \int_{-\pi}^{\pi} d\phi \int_0^R rdr \Big\{1+ \nonumber\\
& \frac{q-1}{T_0} \big[ m_T \cosh(\rho) -p_T \sinh(\rho)
\cos(\phi)\big] \Big\}^{-q/(q-1)},
\end{align}
where $C_2$ is the normalized constant which results in
$\int_0^{\infty} f_2(p_T)dp_T=1$, $q$ is an entropy index
characterizing the degree of non-equilibrium, and $\phi$ denotes
the azimuth. In the case of $n_0=1$ as used in ref. [14], we have
$\beta_T=2\beta_S/(n_0+2)=(2/3)\beta_S$ due to the same flow
profile as in the BGBW model. We would like to point out that the
index $-q/(q-1)$ in Eq. (2) replaced $-1/(q-1)$ in ref. [14] due
to $q$ being very close to 1. In fact, the difference between the
results corresponding to $-q/(q-1)$ and $-1/(q-1)$ are small in
the Tsallis distribution [32].

According to refs. [15, 16], the improved Tsallis distribution in
terms of $p_T$ is
\begin{align}
f_3(p_T)&=C_3 \bigg\{ 2T_0[rI_0(s)K_1(r)-sI_1(s)K_0(r)] \nonumber\\
&-(q-1)T_0r^2I_0(s)[K_0(r)+K_2(r)] \nonumber\\
&+4(q-1)T_0rsI_1(s)K_1(r) \nonumber\\
&-(q-1)T_0s^2K_0(r)[I_0(s)+I_2(s)] \nonumber\\
&+\frac{(q-1)}{4}T_0r^3I_0(s)[K_3(r)+3K_1(r)] \nonumber\\
&-\frac{3(q-1)}{2}T_0r^2s[K_2(r)+K_0(r)]I_1(s) \nonumber\\
&+\frac{3(q-1)}{2}T_0s^2r[I_0(s)+I_2(s)]K_1(r) \nonumber\\
&-\frac{(q-1)}{4}T_0s^3[I_3(s)+3I_1(s)]K_0(r) \bigg\},
\end{align}
where $C_3$ is the normalized constant which results in
$\int_0^{\infty} f_3(p_T)dp_T=1$, $r\equiv \gamma m_T/T_0$,
$s\equiv \gamma \beta_T p_T/T_0$, $\gamma=1/\sqrt{1-\beta_T^2}$,
and $I_{0-3}(s)$ and $K_{0-3}(r)$ are the modified Bessel
functions of the first and second kinds respectively.

As for the alternative method [12, 17--20, 22--24], to use the
relations $T=T_0+am_0$, $\langle p_T \rangle =\langle p_T
\rangle_0+\beta_T \overline{m}$, and  $\langle p \rangle=\langle p
\rangle_0+ \beta \overline{m}$, we can choose the standard and
Tsallis distributions to fit the $p_T$ spectra of identified
particles produced in high energy collisions. Because we give up
the Fermi-Dirac and Bose-Einstein distributions, only the
Boltzmann distribution is used in the case of considering the
standard distribution in the present work. Both the Boltzmann and
Tsallis distributions have more than one forms. We choose the form
of Boltzmann distribution [33]
\begin{align}
f_{4a}(p_T)=C_{4a} p_T m_T \exp \bigg(-\frac{m_T}{T} \bigg)
\end{align}
and the form of Tsallis distribution [32, 33]
\begin{align}
f_{4b}(p_T)=C_{4b} p_T m_T \bigg(1+ \frac{q-1}{T} m_T
\bigg)^{-q/(q-1)},
\end{align}
where $C_{4a}$ and $C_{4b}$ are the normalized constants which
result in $\int_0^{\infty} f_{4a}(p_T)dp_T=1$ and $\int_0^{\infty}
f_{4b}(p_T)dp_T=1$ respectively.

It should be noticed that the above five distributions are only
valid for the spectra in a low-$p_T$ range. That is, they describe
only the soft excitation process. Even if for the soft process,
the Boltzmann distribution is not always enough to fit the $p_T$
spectra in some cases. In fact, two- or three-component Boltzmann
distribution can be used if necessary, in which $T$ is the average
weighted the effective temperatures obtained from different
components. We have
\begin{align}
f_{4a}(p_T)=\sum_{i=1}^{l} k_i C_{4ai} p_T m_T \exp
\bigg(-\frac{m_T}{T_i} \bigg)
\end{align}
and
\begin{align}
T=\sum_{i=1}^{l} k_i T_i,
\end{align}
where $l=2$ or 3 denotes the number of components, and $k_i$,
$C_{4ai}$, and $T_i$ denote the contribution ratio (relative
contribution or fraction), normalization constant, and effective
temperature related to the $i$-th component, respectively. As can
be seen in the next section, Eqs. (6) and (7) are not needed in
the present work due to only simple component Boltzmann
distribution, i.e. Eq. (4), is used in the analyses. We present
here Eqs. (6) and (7) to point out a possible application in
future.

For the spectra in a wide $p_T$ range which contains low and high
$p_T$ regions, we have to consider the contribution of hard
scattering process. Generally, the contribution of hard process is
parameterized to an inverse power-law
\begin{align}
f_H(p_T)=Ap_T \bigg( 1+\frac{p_T}{ p_0} \bigg)^{-n}
\end{align}
which is resulted from the QCD (quantum chromodynamics)
calculation [34--36], where $p_0$ and $n$ are free parameters, and
$A$ is the normalized constant which depends on $p_0$ and $n$ and
results in $\int_0^{\infty} f_H(p_T)dp_T=1$.

To describe the spectra in a wide $p_T$ range, we can use a
superposition of both contributions of soft and hard processes.
The contribution of soft process is described by one of the BGBW
model, the TBW model, the improved Tsallis distribution, the
Boltzmann distribution or two- or three-component Boltzmann
distribution, and the Tsallis distribution, while the contribution
of hard process is described by the inverse power-law. We have the
superposition
\begin{align}
f_0(p_T)=kf_S(p_T)+(1-k)f_H(p_T),
\end{align}
where $k$ denotes the contribution ratio of the soft process and
results naturally in $\int_0^{\infty} f_0(p_T)dp_T=1$, and
$f_S(p_T)$ denotes one of the five distributions discussed in the
four methods.

It should be noted that Eq. (9) and its components $f_S(p_T)$ and
$f_H(p_T)$ are probability density functions. The experimental
quantity of $p_T$ distribution has mainly three forms, $dN/dp_T$,
$d^2N/(dydp_T)$, and $(2\pi p_T)^{-1}d^2N/(dydp_T)$, where $N$
denotes the number of particles and $dy$ is approximately treated
as a constant due to it being usually a given and small value at
the mid-rapidity. To connect Eq. (9) with $dN/dp_T$, we need a
normalization constant $N_0$. To connect Eq. (9) with
$d^2N/(dydp_T)$, we need another normalization constant $N_0$. To
connect Eq. (9) with $(2\pi p_T)^{-1}d^2N/(dydp_T)$, we have to
rewrite Eq. (9) to $f_0(p_T)/p_T=[kf_S(p_T)+(1-k)f_H(p_T)]/p_T$
and compare the right side of the new equation with the data with
a new normalization constant $N_0$.
\\

\begin{figure*}
\hskip-1.0cm
\begin{center}
\includegraphics[width=16.0cm]{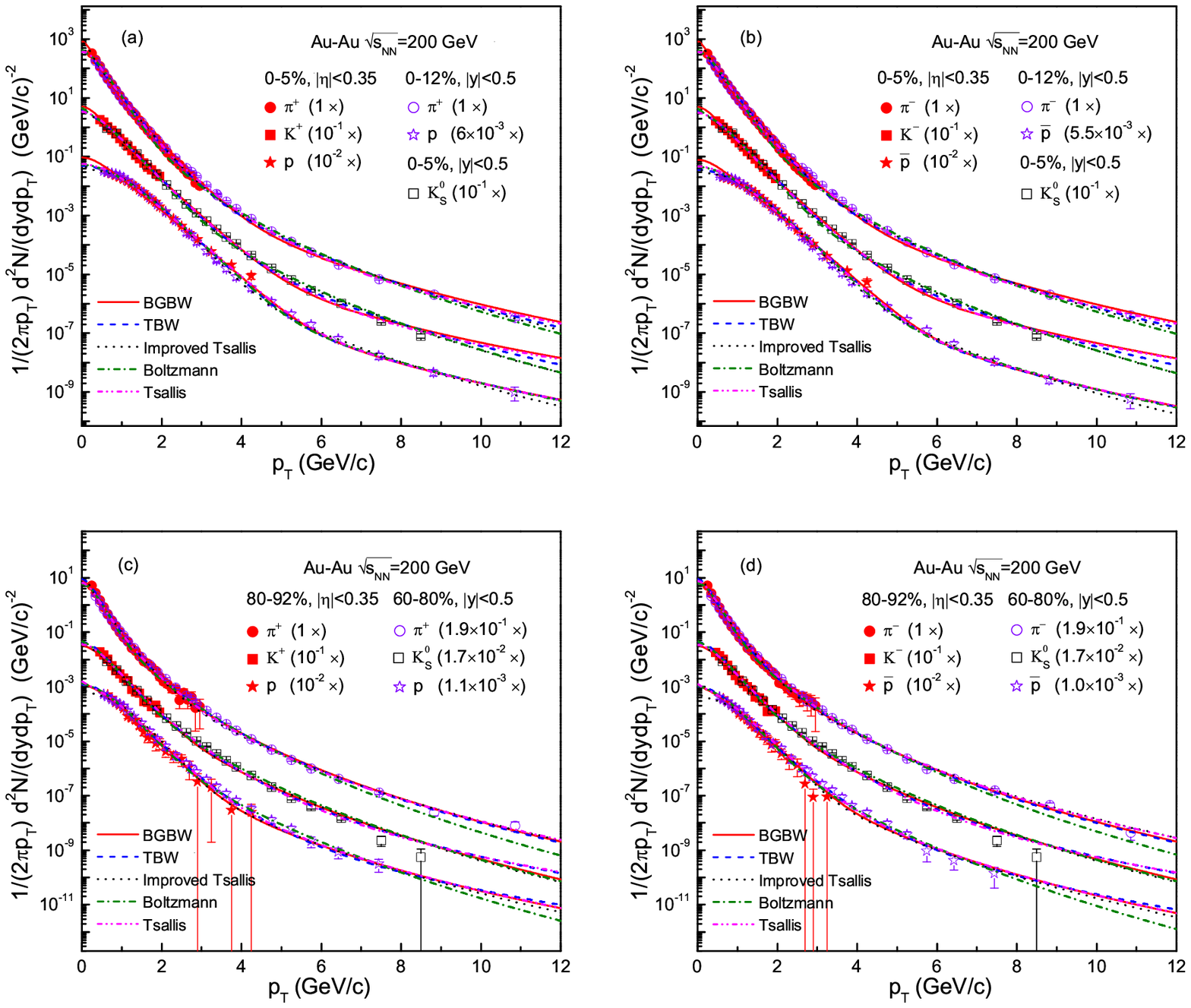}
\end{center}
\vskip0.5cm {\small Fig. 1. (Colour figure online) Transverse
momentum spectra of (a)-(c) $\pi^+$, $K^+$, $K_S^0$, and $p$, as
well as (b)-(d) $\pi^-$, $K^-$, $K_S^0$, and $\bar p$ produced in
(a)-(b) central (0--5\% and 0--12\%) and (c)-(d) peripheral
(80--92\% and 60--80\%) Au-Au collisions at $\sqrt{s_{NN}}=200$
GeV, where the spectra for different types of particles and for
the same or similar particles in different conditions are
multiplied by different amounts shown in the panels for the
clarity and normalization. The closed symbols represent the
experimental data of the PHENIX Collaboration measured in
$|\eta|<0.35$ [25]. The open symbols represent the STAR data
measured in $|y|<0.5$ [26, 27], where the data for $K^+$ and $K^-$
are not available and the data for $K_S^0$ in (a)-(c) and (b)-(d)
are the same. The solid, dashed, dotted, dashed-dotted, and
dashed-dotted-dotted curves are our results calculated by using
the methods i), ii), iii), iv)$_{\rm a}$, and iv)$_{\rm b}$,
respectively.}
\end{figure*}

{\section{Results and discussion}}

Figure 1 presents the transverse momentum spectra, $(2\pi
p_T)^{-1} d^2N/(dydp_T)$, of (a)-(c) positively charged pions
($\pi^+$), positively charged kaons ($K^+$), neutral kaons
($K_S^0$ only), and protons ($p$), as well as (b)-(d) negatively
charged pions ($\pi^-$), negatively charged kaons ($K^-$), neutral
kaons ($K_S^0$ only), and antiprotons ($\bar p$) produced in
(a)-(b) central (0--5\% and 0--12\%) and (c)-(d) peripheral
(80--92\% and 60--80\%) Au-Au collisions at $\sqrt{s_{NN}}=200$
GeV, where the spectra for different types of particles and for
the same or similar particles in different conditions are
multiplied by different amounts shown in the panels for the
clarity and normalization. The closed symbols represent the
experimental data of the PHENIX Collaboration measured in the
pseudorapidity range $|\eta|<0.35$ [25]. The open symbols
represent the STAR data measured in the rapidity range $|y|<0.5$
[26, 27], where the data for $K^+$ and $K^-$ are not available and
the data for $K_S^0$ in (a)-(c) and (b)-(d) are the same. The
solid, dashed, dotted, dashed-dotted, and dashed-dotted-dotted
curves are our results calculated by using the superpositions of
i) the BGBW model (Eq. (1)) and inverse power-law (Eq. (8)), ii)
the TBW model (Eq. (2)) and inverse power-law, iii) the improved
Tsallis distribution (Eq. (3)) and inverse power-law, iv)$_{\rm
a}$ the Boltzmann distribution (Eq. (4)) and inverse power-law, as
well as iv)$_{\rm b}$ the Tsallis distribution (Eq. (5)) and
inverse power-law, respectively. These different superpositions
are also different methods for fitting the data. The values of
free parameters $T_0$, $\beta_T$, $k$, $p_0$, and $n$,
normalization constant $N_0$ which is used to fit the data by a
more accurate method comparing with Ref. [37], and $\chi^2$ per
degree of freedom ($\chi^2$/dof) corresponding to the fit of
method i) are listed in Table 1; the values of $T_0$, $q$,
$\beta_T$, $k$, $p_0$, $n$, $N_0$, and $\chi^2$/dof corresponding
to the methods ii) and iii) are listed in Tables 2 and 3
respectively; the values of $T$, $k$, $p_0$, $n$, $N_0$, and
$\chi^2$/dof corresponding to the methods iv)$_{\rm a}$ are listed
in Table 4; and the values of $T$, $q$, $k$, $p_0$, $n$, $N_0$,
and $\chi^2$/dof corresponding to the methods iv)$_{\rm b}$ are
listed in Table 5. One can see that, in most cases, all of the
considered methods describe approximately the $p_T$ spectra of
identified particles produced in central and peripheral Au-Au
collisions at $\sqrt{s_{NN}}=200$ GeV.

\begin{figure*}
\hskip-1.0cm \begin{center}
\includegraphics[width=16.0cm]{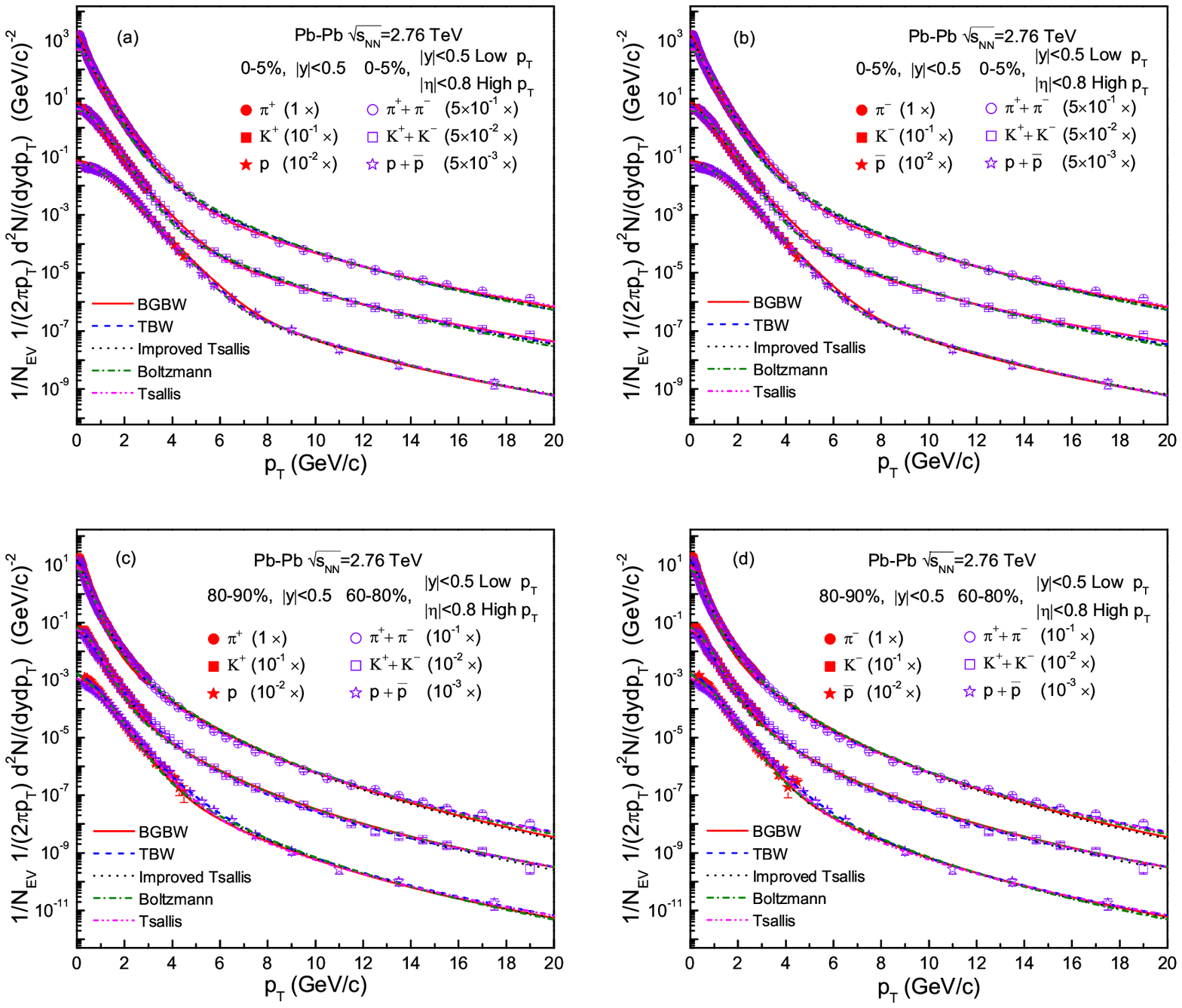}
\end{center}
\vskip0.5cm {\small Fig. 2. (Colour figure online) Same as Fig. 1,
but showing the spectra of (a)-(c) $\pi^+$ ($\pi^++\pi^-$), $K^+$
($K^++K^-$), and $p$ ($p+\bar p$), as well as (b)-(d) $\pi^-$
($\pi^++\pi^-$), $K^-$ ($K^++K^-$), $\bar p$ ($p+\bar p$) produced
in (a)-(b) central (0--5\%) and (c)-(d) peripheral (80--90\% and
60--80\%) Pb-Pb collisions at $\sqrt{s_{NN}}=2.76$ TeV, where
$N_{EV}$ on the vertical axis denotes the number of events, which
is usually omitted. The closed (open) symbols represent the
experimental data of the ALICE Collaboration measured in $|y|<0.5$
[28] (in $|\eta|<0.8$ for high $p_T$ region and in $|y|<0.5$ for
low $p_T$ region [29]). The data for $\pi^++\pi^-$, $K^++K^-$, and
$p+\bar p$ in (a)-(c) and (b)-(d) are the same.}
\end{figure*}

\begin{table*}
{\scriptsize Table 1. Values of free parameters ($T_0$,
$\beta_{T}$, $k$, $p_{0}$, and $n$), normalization constant
($N_{0}$), and $\chi^2$/dof corresponding to the fits of method i)
in Figs. 1 and 2, where $n_0=2$ in the self-similar flow profile
is used as ref. [11].
\begin{center}
\begin{tabular}{cccccccccc}
\hline Fig. & Cent. & Main Part. & $T_0$ (GeV) & $\beta_{T}$ ($c$) & $k$ & $p_{0}$ (GeV/$c$) & $n$ & $N_0$ & $\chi^2$/dof \\
\hline
1(a) &  Central  & $\pi^+$   & $0.113\pm0.004$ & $0.413\pm0.006$ & $0.988\pm0.003$ & $2.075\pm0.062$ & $9.015\pm0.148$ & $985.309\pm75.105$  & $2.708$\\
     &           & $K^+$     & $0.124\pm0.005$ & $0.408\pm0.006$ & $0.975\pm0.004$ & $1.295\pm0.058$ & $7.375\pm0.128$ & $65.180\pm3.873$    & $7.300$\\
     &           & $p$       & $0.127\pm0.005$ & $0.392\pm0.006$ & $0.989\pm0.003$ & $2.485\pm0.076$ & $8.775\pm0.136$ & $12.611\pm0.636$    & $10.729$\\
1(b) &  Central  & $\pi^-$   & $0.113\pm0.004$ & $0.413\pm0.006$ & $0.988\pm0.003$ & $2.075\pm0.062$ & $9.015\pm0.148$ & $985.309\pm75.105$  & $2.792$\\
     &           & $K^-$     & $0.124\pm0.005$ & $0.408\pm0.006$ & $0.975\pm0.004$ & $1.295\pm0.058$ & $7.375\pm0.128$ & $63.109\pm3.589$    & $9.267$\\
     &           & $\bar{p}$ & $0.125\pm0.005$ & $0.392\pm0.006$ & $0.990\pm0.003$ & $2.465\pm0.076$ & $8.895\pm0.136$ & $10.572\pm0.599$    & $20.512$\\
1(c) & Peripheral& $\pi^+$   & $0.160\pm0.004$ & $0.000^{+0.016}_{-0}$ & $0.754\pm0.009$ & $2.012\pm0.069$ & $10.803\pm0.143$ & $9.794\pm0.726$ & $4.237$\\
     &           & $K^+$     & $0.238\pm0.005$ & $0.000^{+0.016}_{-0}$ & $0.825\pm0.009$ & $3.383\pm0.089$ & $12.313\pm0.166$ & $0.424\pm0.029$ & $8.755$\\
     &           & $p$       & $0.251\pm0.005$ & $0.000^{+0.016}_{-0}$ & $0.851\pm0.011$ & $2.006\pm0.065$ & $9.466\pm0.139$  & $0.167\pm0.012$ & $1.434$\\
1(d) & Peripheral& $\pi^-$   & $0.160\pm0.004$ & $0.000^{+0.016}_{-0}$ & $0.754\pm0.009$ & $2.012\pm0.069$ & $10.803\pm0.143$ & $9.794\pm0.726$ & $3.792$\\
     &           & $K^-$     & $0.238\pm0.005$ & $0.000^{+0.016}_{-0}$ & $0.825\pm0.009$ & $3.383\pm0.089$ & $12.313\pm0.166$ & $0.424\pm0.029$ & $7.469$\\
     &           & $\bar{p}$ & $0.251\pm0.005$ & $0.000^{+0.016}_{-0}$ & $0.851\pm0.011$ & $2.106\pm0.066$ & $9.766\pm0.141$  & $0.134\pm0.007$ & $0.834$\\
\hline
2(a) &  Central  & $\pi^+$   & $0.128\pm0.004$ & $0.434\pm0.007$ & $0.992\pm0.002$ & $2.775\pm0.091$ & $7.435\pm0.133$ & $1771.569\pm112.825$ & $1.200$\\
     &           & $K^+$     & $0.187\pm0.004$ & $0.390\pm0.006$ & $0.993\pm0.002$ & $3.575\pm0.098$ & $7.135\pm0.128$ & $92.874\pm6.429$     & $3.647$\\
     &           & $p$       & $0.429\pm0.005$ & $0.145\pm0.005$ & $0.976\pm0.005$ & $2.485\pm0.088$ & $7.375\pm0.136$ & $10.188\pm0.445$     & $7.472$\\
2(b) &  Central  & $\pi^-$   & $0.128\pm0.004$ & $0.434\pm0.007$ & $0.992\pm0.002$ & $2.775\pm0.091$ & $7.435\pm0.133$ & $1771.569\pm112.825$ & $1.221$\\
     &           & $K^-$     & $0.187\pm0.004$ & $0.390\pm0.006$ & $0.993\pm0.002$ & $3.575\pm0.098$ & $7.135\pm0.128$ & $92.874\pm6.429$     & $3.288$\\
     &           & $\bar{p}$ & $0.428\pm0.005$ & $0.145\pm0.005$ & $0.976\pm0.005$ & $2.485\pm0.088$ & $7.375\pm0.136$ & $10.209\pm0.446$     & $6.875$\\
2(c) & Peripheral& $\pi^+$   & $0.183\pm0.004$ & $0.000^{+0.017}_{-0}$ & $0.909\pm0.009$ &$2.793\pm0.089$ &$8.985\pm0.133$ &$12.144\pm0.705$  & $16.627$\\
     &           & $K^+$     & $0.272\pm0.004$ & $0.000^{+0.017}_{-0}$ & $0.835\pm0.009$ &$2.375\pm0.085$ &$7.885\pm0.165$ &$0.707\pm0.028$   & $2.808$\\
     &           & $p$       & $0.338\pm0.004$ & $0.000^{+0.017}_{-0}$ & $0.836\pm0.009$ &$1.875\pm0.078$ &$7.705\pm0.138$ &$0.183\pm0.011$   & $2.752$\\
2(d) & Peripheral& $\pi^-$   & $0.183\pm0.004$ & $0.000^{+0.017}_{-0}$ & $0.909\pm0.009$ &$2.793\pm0.089$ &$8.985\pm0.133$ &$12.144\pm0.705$  & $16.734$\\
     &           & $K^-$     & $0.272\pm0.004$ & $0.000^{+0.017}_{-0}$ & $0.835\pm0.009$ &$2.375\pm0.085$ &$7.885\pm0.165$ &$0.707\pm0.028$   & $3.041$\\
     &           & $\bar{p}$ & $0.342\pm0.004$ & $0.000^{+0.017}_{-0}$ & $0.815\pm0.009$ &$1.875\pm0.078$ &$7.705\pm0.138$ &$0.185\pm0.012$   & $2.602$\\
\hline
\end{tabular}
\end{center}}
\end{table*}

\begin{table*}
{\tiny Table 2. Values of free parameters ($T_0$, $q$,
$\beta_{T}$, $k$, $\ p_{0}$, and $n$), normalization constant
($N_{0}$), and $\chi^2$/dof corresponding to the fits of method
ii) in Figs. 1 and 2, where $n_0=1$ is used as ref. [14].
\begin{center}
\begin{tabular}{ccccccccccc}
\hline Fig. & Cent. & Main Part. & $T_0$ (GeV) & $q$ & $\beta_{T}$ ($c$) & $k$ & $p_{0}$ (GeV/$c$) & $n$ & $N_0$ & $\chi^2$/dof \\
\hline
1(a) &  Central  & $\pi^+$   & $0.108\pm0.004$ & $1.008\pm0.005$ & $0.472\pm0.010$ & $0.882\pm0.008$ & $1.775\pm0.069$ & $9.895\pm0.143$ & $486.350\pm40.221$ & $4.082$ \\
     &           & $K^+$     & $0.113\pm0.004$ & $1.020\pm0.005$ & $0.469\pm0.010$ & $0.901\pm0.006$ & $1.875\pm0.072$ & $9.405\pm0.139$ & $44.575\pm2.808$   & $6.564$ \\
     &           & $p$       & $0.119\pm0.004$ & $1.011\pm0.004$ & $0.469\pm0.008$ & $0.989\pm0.003$ & $2.885\pm0.082$ & $9.275\pm0.136$ & $7.214\pm0.368$    & $1.665$ \\
1(b) &  Central  & $\pi^-$   & $0.108\pm0.004$ & $1.008\pm0.005$ & $0.472\pm0.010$ & $0.882\pm0.008$ & $1.775\pm0.069$ & $9.895\pm0.143$ & $486.350\pm40.221$ & $3.856$ \\
     &           & $K^-$     & $0.113\pm0.004$ & $1.020\pm0.005$ & $0.469\pm0.010$ & $0.901\pm0.006$ & $1.875\pm0.072$ & $9.405\pm0.139$ & $42.837\pm2.808$   & $5.939$ \\
     &           & $\bar{p}$ & $0.121\pm0.004$ & $1.010\pm0.004$ & $0.469\pm0.008$ & $0.991\pm0.003$ & $2.885\pm0.082$ & $9.305\pm0.134$ & $5.369\pm0.354$    & $6.643$ \\
1(c) & Peripheral& $\pi^+$   & $0.099\pm0.004$ & $1.078\pm0.005$ & $0.000^{+0.036}_{-0}$ & $0.862\pm0.008$ & $2.198\pm0.089$ & $10.982\pm0.161$ & $11.341\pm0.747$ & $3.221$ \\
     &           & $K^+$     & $0.119\pm0.004$ & $1.088\pm0.004$ & $0.000^{+0.036}_{-0}$ & $0.985\pm0.008$ & $1.983\pm0.078$ & $8.253\pm0.138$  & $0.589\pm0.053$  & $4.002$ \\
     &           & $p$       & $0.132\pm0.004$ & $1.064\pm0.005$ & $0.000^{+0.036}_{-0}$ & $0.985\pm0.004$ & $2.010\pm0.088$ & $7.966\pm0.129$  & $0.171\pm0.014$  & $0.940$ \\
1(d) & Peripheral& $\pi^-$   & $0.099\pm0.004$ & $1.078\pm0.005$ & $0.000^{+0.036}_{-0}$ & $0.862\pm0.008$ & $2.198\pm0.089$ & $10.982\pm0.161$ & $11.341\pm0.747$ & $2.924$ \\
     &           & $K^-$     & $0.119\pm0.004$ & $1.088\pm0.004$ & $0.000^{+0.036}_{-0}$ & $0.985\pm0.008$ & $1.983\pm0.078$ & $8.253\pm0.138$  & $0.589\pm0.053$  & $3.652$ \\
     &           & $\bar{p}$ & $0.124\pm0.004$ & $1.067\pm0.005$ & $0.000^{+0.036}_{-0}$ & $0.983\pm0.004$ & $2.018\pm0.088$ & $8.166\pm0.129$  & $0.144\pm0.014$  & $0.552$ \\
\hline
2(a) &  Central  & $\pi^+$   & $0.109\pm0.004$ & $1.009\pm0.005$ & $0.525\pm0.009$ & $0.977\pm0.005$ & $2.585\pm0.086$ & $7.875\pm0.122$ & $917.576\pm91.809$  & $6.313$ \\
     &           & $K^+$     & $0.145\pm0.005$ & $1.004\pm0.003$ & $0.500\pm0.009$ & $0.984\pm0.004$ & $3.255\pm0.091$ & $7.508\pm0.119$ & $66.904\pm6.743$    & $0.580$ \\
     &           & $p$       & $0.178\pm0.005$ & $1.002\pm0.001$ & $0.500\pm0.008$ & $0.993\pm0.002$ & $4.975\pm0.099$ & $8.725\pm0.121$ & $8.981\pm0.254$     & $3.509$ \\
2(b) &  Central  &$\pi^-$    & $0.109\pm0.004$ & $1.009\pm0.005$ & $0.525\pm0.009$ & $0.977\pm0.005$ & $2.585\pm0.086$ & $7.875\pm0.122$ & $917.576\pm91.809$  & $6.249$ \\
     &           &$K^-$      & $0.145\pm0.005$ & $1.004\pm0.003$ & $0.500\pm0.009$ & $0.985\pm0.004$ & $3.255\pm0.091$ & $7.508\pm0.119$ & $66.904\pm6.743$    & $0.570$ \\
     &           &$\bar{p}$  & $0.178\pm0.005$ & $1.002\pm0.001$ & $0.500\pm0.008$ & $0.993\pm0.002$ & $4.975\pm0.099$ & $8.725\pm0.121$ & $8.981\pm0.254$     & $3.253$ \\
2(c) & Peripheral& $\pi^+$   & $0.102\pm0.004$ & $1.108\pm0.005$ & $0.000^{+0.018}_{-0}$ & $0.976\pm0.005$ & $3.003\pm0.089$ & $8.335\pm0.118$ & $15.628\pm0.563$ & $10.532$ \\
     &           & $K^+$     & $0.141\pm0.005$ & $1.099\pm0.005$ & $0.000^{+0.018}_{-0}$ & $0.906\pm0.005$ & $1.875\pm0.071$ & $7.038\pm0.109$ & $0.820\pm0.063$  & $1.149$ \\
     &           & $p$       & $0.172\pm0.005$ & $1.076\pm0.005$ & $0.000^{+0.018}_{-0}$ & $0.958\pm0.005$ & $2.375\pm0.088$ & $7.575\pm0.119$ & $0.212\pm0.017$  & $4.623$ \\
2(d) & Peripheral& $\pi^-$   & $0.102\pm0.004$ & $1.108\pm0.005$ & $0.000^{+0.018}_{-0}$ & $0.976\pm0.005$ & $3.003\pm0.089$ & $8.335\pm0.118$ & $15.628\pm0.563$ & $10.481$ \\
     &           & $K^-$     & $0.141\pm0.005$ & $1.099\pm0.005$ & $0.000^{+0.018}_{-0}$ & $0.906\pm0.005$ & $1.875\pm0.071$ & $7.038\pm0.109$ & $0.820\pm0.063$  & $1.279$ \\
     &           & $\bar{p}$ & $0.172\pm0.005$ & $1.076\pm0.005$ & $0.000^{+0.018}_{-0}$ & $0.958\pm0.005$ & $2.375\pm0.088$ & $7.575\pm0.119$ & $0.212\pm0.017$  & $4.832$ \\
\hline
\end{tabular}
\end{center}}
\end{table*}

\begin{table*}
{\tiny Table 3. Values of free parameters ($T_0$, $q$,
$\beta_{T}$, $k$, $p_{0}$, and $n$), normalization constant
($N_{0}$), and $\chi^2$/dof corresponding to the fits of method
iii) in Figs. 1 and 2.
\begin{center}
\begin{tabular}{ccccccccccc}
\hline Fig. & Cent. & Main Part. & $T_0$ (GeV) & $q$ & $\beta_{T}$ ($c$) & $k$ & $p_{0}$ (GeV/$c$) & $n$ & $N_0$ & $\chi^2$/dof \\
\hline
1(a) &  Central  & $\pi^+$   & $0.113\pm0.006$ & $1.017\pm0.007$ & $0.634\pm0.009$ & $0.939\pm0.008$ & $2.475\pm0.088$ & $11.091\pm0.165$ & $746.564\pm89.743$  & $2.986$ \\
     &           & $K^+$     & $0.116\pm0.006$ & $1.040\pm0.007$ & $0.634\pm0.009$ & $0.902\pm0.008$ & $3.675\pm0.091$ & $12.995\pm0.172$ & $32.457\pm5.734$    & $9.781$ \\
     &           & $p$       & $0.121\pm0.006$ & $1.024\pm0.007$ & $0.634\pm0.009$ & $0.916\pm0.008$ & $2.985\pm0.090$ & $11.225\pm0.162$ & $5.365\pm0.677$     & $1.249$ \\
1(b) &  Central  & $\pi^-$   & $0.113\pm0.006$ & $1.017\pm0.007$ & $0.634\pm0.009$ & $0.939\pm0.008$ & $2.475\pm0.088$ & $11.091\pm0.165$ & $746.564\pm89.743$  & $2.700$ \\
     &           & $K^-$     & $0.116\pm0.006$ & $1.040\pm0.007$ & $0.634\pm0.009$ & $0.900\pm0.008$ & $3.675\pm0.091$ & $12.995\pm0.172$ & $31.193\pm5.698$    & $8.100$ \\
     &           & $\bar{p}$ & $0.121\pm0.006$ & $1.024\pm0.007$ & $0.634\pm0.009$ & $0.909\pm0.008$ & $2.985\pm0.090$ & $11.525\pm0.162$ & $8.294\pm1.243$     & $2.878$ \\
1(c) & Peripheral& $\pi^+$   & $0.102\pm0.006$ & $1.031\pm0.007$ & $0.583\pm0.009$ & $0.891\pm0.008$ & $2.185\pm0.086$ & $10.632\pm0.148$ & $10.292\pm1.860$    & $3.931$ \\
     &           & $K^+$     & $0.109\pm0.006$ & $1.045\pm0.008$ & $0.578\pm0.009$ & $0.872\pm0.008$ & $4.483\pm0.099$ & $14.061\pm0.165$ & $0.327\pm0.057$     & $8.529$ \\
     &           & $p$       & $0.110\pm0.006$ & $1.053\pm0.008$ & $0.548\pm0.008$ & $0.901\pm0.008$ & $3.066\pm0.095$ & $11.166\pm0.126$ & $0.083\pm0.005$     & $2.700$ \\
1(d) & Peripheral& $\pi^-$   & $0.102\pm0.006$ & $1.031\pm0.007$ & $0.583\pm0.009$ & $0.891\pm0.008$ & $2.185\pm0.086$ & $10.532\pm0.148$ & $10.771\pm1.863$    & $3.751$ \\
     &           & $K^-$     & $0.109\pm0.006$ & $1.045\pm0.008$ & $0.578\pm0.009$ & $0.872\pm0.008$ & $4.483\pm0.099$ & $14.061\pm0.165$ & $0.327\pm0.057$     & $7.157$ \\
     &           & $\bar{p}$ & $0.110\pm0.006$ & $1.053\pm0.008$ & $0.548\pm0.008$ & $0.901\pm0.008$ & $3.066\pm0.095$ & $11.166\pm0.126$ & $0.055\pm0.005$     & $1.316$ \\
\hline
2(a) &  Central  & $\pi^+$   & $0.152\pm0.004$ & $1.011\pm0.004$ & $0.609\pm0.010$ & $0.981\pm0.007$ & $2.575\pm0.094$ & $7.775\pm0.145$  & $1475.441\pm93.801$ & $2.682$ \\
     &           & $K^+$     & $0.158\pm0.004$ & $1.059\pm0.008$ & $0.609\pm0.010$ & $0.987\pm0.006$ & $3.575\pm0.102$ & $7.655\pm0.144$  & $58.904\pm5.207$    & $1.235$ \\
     &           & $p$       & $0.194\pm0.005$ & $1.069\pm0.011$ & $0.609\pm0.010$ & $0.987\pm0.006$ & $2.885\pm0.101$ & $7.375\pm0.148$  & $7.792\pm0.559$     & $4.833$ \\
2(b) &  Central  & $\pi^-$   & $0.152\pm0.004$ & $1.011\pm0.004$ & $0.609\pm0.010$ & $0.981\pm0.007$ & $2.575\pm0.094$ & $7.775\pm0.145$  & $1475.441\pm93.801$ & $2.586$ \\
     &           &$K^-$      & $0.158\pm0.004$ & $1.059\pm0.008$ & $0.609\pm0.010$ & $0.987\pm0.006$ & $3.575\pm0.102$ & $7.655\pm0.144$  & $58.904\pm5.207$    & $1.083$ \\
     &           & $\bar{p}$ & $0.194\pm0.005$ & $1.069\pm0.011$ & $0.609\pm0.010$ & $0.987\pm0.006$ & $2.885\pm0.101$ & $7.375\pm0.148$  & $7.792\pm0.559$     & $4.482$ \\
2(c) & Peripheral& $\pi^+$   & $0.118\pm0.005$ & $1.008\pm0.005$ & $0.630\pm0.009$ & $0.920\pm0.007$ & $2.903\pm0.103$ & $9.135\pm0.165$  & $15.956\pm0.981$    & $5.202$ \\
     &           & $K^+$     & $0.143\pm0.004$ & $1.011\pm0.005$ & $0.602\pm0.009$ & $0.901\pm0.007$ & $3.003\pm0.111$ & $8.335\pm0.155$  & $0.530\pm0.038$     & $1.880$ \\
     &           & $p$       & $0.163\pm0.005$ & $1.021\pm0.005$ & $0.559\pm0.009$ & $0.889\pm0.007$ & $2.375\pm0.099$ & $8.059\pm0.142$  & $0.102\pm0.006$     & $2.804$ \\
2(d) & Peripheral& $\pi^-$   & $0.118\pm0.005$ & $1.008\pm0.005$ & $0.630\pm0.009$ & $0.920\pm0.007$ & $2.903\pm0.103$ & $9.135\pm0.165$  & $15.956\pm0.981$    & $5.257$ \\
     &           & $K^-$     & $0.143\pm0.004$ & $1.011\pm0.005$ & $0.602\pm0.009$ & $0.901\pm0.007$ & $3.003\pm0.111$ & $8.335\pm0.155$  & $0.525\pm0.034$     & $1.979$ \\
     &           & $\bar{p}$ & $0.163\pm0.005$ & $1.021\pm0.005$ & $0.559\pm0.009$ & $0.889\pm0.007$ & $2.375\pm0.099$ & $8.059\pm0.142$  & $0.101\pm0.006$     & $2.942$ \\
\hline
\end{tabular}
\end{center}}
\end{table*}

\begin{table*}
{\scriptsize Table 4. Values of free parameters ($T$, $k$,
$p_{0}$, and $n$ ), normalization constant ($N_{0}$), and
$\chi^2$/dof corresponding to the fits of method iv)$_{\rm a}$ in
Figs. 1 and 2.
\begin{center}
\begin{tabular}{ccccccccc}
\hline Fig. & Cent. & Main Part. & $T$ (GeV) & $k$ & $p_{0}$ (GeV/$c$) & $n$ & $N_0$ & $\chi^2$/dof \\
\hline
1(a) &  Central  & $\pi^+$   & $0.167\pm0.004$ & $0.765\pm0.008$ & $2.095\pm0.068$ & $11.295\pm0.133$ & $519.268\pm39.582$& $9.637$ \\
     &           & $K^+$     & $0.235\pm0.004$ & $0.752\pm0.008$ & $2.915\pm0.068$ & $12.335\pm0.185$ & $49.650\pm2.890$  & $12.847$ \\
     &           & $p$       & $0.302\pm0.005$ & $0.983\pm0.005$ & $2.785\pm0.066$ & $9.475\pm0.176$  & $7.744\pm0.267$   & $2.217$ \\
1(b) &  Central  &$\pi^-$    & $0.167\pm0.004$ & $0.765\pm0.008$ & $2.095\pm0.068$ & $11.295\pm0.133$ & $519.297\pm39.582$& $9.068$ \\
     &           &$K^-$      & $0.235\pm0.004$ & $0.750\pm0.008$ & $2.915\pm0.068$ & $12.335\pm0.185$ & $47.297\pm2.893$  & $13.624$ \\
     &           & $\bar{p}$ & $0.296\pm0.005$ & $0.981\pm0.005$ & $2.715\pm0.066$ & $9.675\pm0.176$  & $6.516\pm0.272$   & $6.399$ \\
1(c) & Peripheral& $\pi^+$   & $0.131\pm0.004$ & $0.799\pm0.008$ & $3.238\pm0.089$ & $13.892\pm0.132$ & $8.602\pm0.676$   & $4.243$ \\
     &           & $K^+$     & $0.185\pm0.004$ & $0.702\pm0.008$ & $3.483\pm0.086$ & $13.083\pm0.146$ & $0.556\pm0.035$   & $6.799$ \\
     &           & $p$       & $0.209\pm0.005$ & $0.822\pm0.008$ & $4.606\pm0.106$ & $14.866\pm0.155$ & $0.173\pm0.012$   & $0.955$ \\
1(d) & Peripheral& $\pi^-$   & $0.131\pm0.004$ & $0.799\pm0.008$ & $3.238\pm0.089$ & $13.892\pm0.132$ & $8.602\pm0.676$   & $4.115$ \\
     &           & $K^-$     & $0.185\pm0.004$ & $0.702\pm0.008$ & $3.483\pm0.086$ & $13.083\pm0.146$ & $0.559\pm0.035$   & $6.284$ \\
     &           & $\bar{p}$ & $0.209\pm0.005$ & $0.822\pm0.008$ & $4.606\pm0.106$ & $15.279\pm0.165$ & $0.139\pm0.012$   & $0.627$ \\
\hline
2(a) &  Central  & $\pi^+$   & $0.215\pm0.004$ & $0.828\pm0.008$ & $1.375\pm0.068$ & $7.315\pm0.128$  & $679.491\pm44.189$& $16.706$ \\
     &           & $K^+$     & $0.299\pm0.005$ & $0.972\pm0.008$ & $2.945\pm0.090$ & $7.685\pm0.132$  & $57.722\pm5.536$  & $1.889$ \\
     &           & $p$       & $0.413\pm0.005$ & $0.993\pm0.002$ & $4.975\pm0.112$ & $8.725\pm0.146$  & $8.864\pm0.467$   & $2.600$ \\
2(b) &  Central  &$\pi^-$    & $0.215\pm0.004$ & $0.828\pm0.008$ & $1.375\pm0.068$ & $7.315\pm0.128$  & $679.491\pm44.189$& $16.821$ \\
     &           &$K^-$      & $0.299\pm0.005$ & $0.972\pm0.008$ & $2.945\pm0.090$ & $7.685\pm0.132$  & $57.722\pm5.536$  & $2.052$ \\
     &           &$\bar{p}$  & $0.413\pm0.005$ & $0.993\pm0.002$ & $4.975\pm0.112$ & $8.725\pm0.146$  & $8.864\pm0.467$   & $2.433$ \\
2(c) & Peripheral& $\pi^+$   & $0.152\pm0.004$ & $0.802\pm0.008$ & $2.012\pm0.065$ & $8.279\pm0.116$  & $9.713\pm0.616$   & $15.656$ \\
     &           & $K^+$     & $0.219\pm0.004$ & $0.803\pm0.009$ & $2.035\pm0.092$ & $7.595\pm0.134$  & $0.822\pm0.052$   & $5.123$ \\
     &           & $p$       & $0.291\pm0.005$ & $0.805\pm0.008$ & $2.285\pm0.096$ & $8.365\pm0.142$  & $0.190\pm0.017$   & $3.545$ \\
2(d) & Peripheral& $\pi^-$   & $0.152\pm0.004$ & $0.802\pm0.008$ & $2.012\pm0.065$ & $8.279\pm0.116$  & $9.713\pm0.616$   & $15.657$ \\
     &           & $K^-$     & $0.219\pm0.004$ & $0.803\pm0.009$ & $2.035\pm0.092$ & $7.595\pm0.134$  & $0.822\pm0.052$   & $5.238$ \\
     &           & $\bar{p}$ & $0.296\pm0.005$ & $0.805\pm0.008$ & $2.285\pm0.096$ & $8.365\pm0.142$  & $0.188\pm0.017$   & $3.391$ \\
\hline
\end{tabular}
\end{center}}
\end{table*}

\begin{table*}
{\scriptsize Table 5. Values of free parameters ($T$, $q$, $k$,
$p_{0}$, and $n$), normalization constant ($N_{0}$), and
$\chi^2$/dof corresponding to the fits of method iv)$_{\rm b}$ in
Figs. 1 and 2.
\begin{center}
\begin{tabular}{cccccccccc}
\hline Fig. & Cent. & Main Part. & $T$ (GeV) & $q$ & $k$ & $p_{0}$ (GeV/$c$) & $n$ & $N_0$ & $\chi^2$/dof \\
\hline
1(a) &  Central  & $\pi^+$   & $0.130\pm0.004$ & $1.073\pm0.003$ & $0.994\pm0.003$ & $1.775\pm0.069$ & $8.115\pm0.148$ & $508.830\pm43.650$& $1.731$ \\
     &           & $K^+$     & $0.184\pm0.005$ & $1.050\pm0.004$ & $0.984\pm0.005$ & $1.075\pm0.058$ & $6.775\pm0.135$ & $45.687\pm2.962$  & $4.354$ \\
     &           & $p$       & $0.274\pm0.004$ & $1.015\pm0.003$ & $0.988\pm0.003$ & $2.485\pm0.088$ & $8.775\pm0.152$ & $8.211\pm0.194$   & $3.268$ \\
1(b) &  Central  &$\pi^-$    & $0.130\pm0.004$ & $1.073\pm0.003$ & $0.994\pm0.003$ & $1.775\pm0.069$ & $8.115\pm0.148$ & $508.830\pm43.650$& $1.648$ \\
     &           &$K^-$      & $0.184\pm0.005$ & $1.050\pm0.004$ & $0.982\pm0.005$ & $1.075\pm0.058$ & $6.775\pm0.135$ & $42.366\pm2.868$  & $2.951$ \\
     &           & $\bar{p}$ & $0.272\pm0.004$ & $1.012\pm0.003$ & $0.992\pm0.003$ & $2.985\pm0.090$ & $9.375\pm0.159$ & $6.764\pm0.189$   & $7.806$ \\
1(c) & Peripheral& $\pi^+$   & $0.105\pm0.004$ & $1.085\pm0.005$ & $0.918\pm0.005$ & $1.985\pm0.075$ & $10.032\pm0.155$& $8.344\pm0.606$   & $1.855$ \\
     &           & $K^+$     & $0.137\pm0.004$ & $1.079\pm0.004$ & $0.990\pm0.006$ & $1.983\pm0.075$ & $7.853\pm0.136$ & $0.488\pm0.033$   & $3.574$ \\
     &           & $p$       & $0.192\pm0.005$ & $1.028\pm0.006$ & $0.853\pm0.008$ & $2.006\pm0.056$ & $9.466\pm0.155$ & $0.175\pm0.012$   & $1.165$ \\
1(d) & Peripheral& $\pi^-$   & $0.105\pm0.004$ & $1.085\pm0.005$ & $0.918\pm0.005$ & $1.985\pm0.075$ & $10.032\pm0.155$& $8.344\pm0.606$   & $1.635$ \\
     &           & $K^-$     & $0.137\pm0.004$ & $1.079\pm0.004$ & $0.990\pm0.006$ & $1.983\pm0.075$ & $7.853\pm0.136$ & $0.466\pm0.030$   & $2.604$ \\
     &           & $\bar{p}$ & $0.192\pm0.005$ & $1.028\pm0.006$ & $0.853\pm0.008$ & $2.106\pm0.059$ & $9.766\pm0.158$ & $0.140\pm0.012$   & $0.715$ \\
\hline
2(a) &  Central  & $\pi^+$   & $0.170\pm0.005$ & $1.066\pm0.005$ & $0.992\pm0.007$ & $2.775\pm0.062$ & $7.275\pm0.185$ & $711.631\pm55.063$ & $6.847$ \\
     &           & $K^+$     & $0.264\pm0.006$ & $1.030\pm0.005$ & $0.993\pm0.002$ & $3.575\pm0.108$ & $7.135\pm0.203$ & $62.036\pm5.422$   & $0.548$ \\
     &           & $p$       & $0.409\pm0.006$ & $1.002\pm0.001$ & $0.993\pm0.002$ & $4.975\pm0.112$ & $8.725\pm0.206$ & $8.968\pm0.417$    & $2.813$ \\
2(b) &  Central  &$\pi^-$    & $0.170\pm0.005$ & $1.066\pm0.005$ & $0.992\pm0.007$ & $2.775\pm0.062$ & $7.275\pm0.185$ & $711.631\pm55.063$ & $6.813$ \\
     &           &$K^-$      & $0.264\pm0.006$ & $1.030\pm0.005$ & $0.993\pm0.002$ & $3.575\pm0.108$ & $7.135\pm0.203$ & $62.036\pm5.422$   & $0.654$ \\
     &           & $\bar{p}$ & $0.409\pm0.006$ & $1.002\pm0.001$ & $0.993\pm0.002$ & $4.975\pm0.112$ & $8.725\pm0.206$ & $8.968\pm0.417$    & $2.651$ \\
2(c) & Peripheral& $\pi^+$   & $0.117\pm0.004$ & $1.099\pm0.005$ & $0.972\pm0.005$ & $3.003\pm0.098$ & $8.335\pm0.196$ & $10.635\pm0.595$   & $7.995$ \\
     &           & $K^+$     & $0.173\pm0.005$ & $1.069\pm0.005$ & $0.905\pm0.006$ & $2.375\pm0.071$ & $7.575\pm0.192$ & $0.725\pm0.043$    & $1.674$ \\
     &           & $p$       & $0.263\pm0.005$ & $1.035\pm0.005$ & $0.911\pm0.006$ & $1.875\pm0.065$ & $7.265\pm0.146$ & $0.139\pm0.009$    & $2.285$ \\
2(d) & Peripheral& $\pi^-$   & $0.117\pm0.004$ & $1.099\pm0.005$ & $0.972\pm0.005$ & $3.003\pm0.098$ & $8.335\pm0.196$ & $10.635\pm0.595$   & $7.904$ \\
     &           & $K^-$     & $0.173\pm0.005$ & $1.069\pm0.005$ & $0.905\pm0.006$ & $2.375\pm0.071$ & $7.575\pm0.192$ & $0.725\pm0.043$    & $1.875$ \\
     &           & $\bar{p}$ & $0.263\pm0.005$ & $1.035\pm0.005$ & $0.911\pm0.006$ & $1.875\pm0.065$ & $7.265\pm0.146$ & $0.144\pm0.009$    & $2.255$ \\
\hline
\end{tabular}
\end{center}}
\end{table*}

\begin{figure*}
\hskip-1.0cm \begin{center}
\includegraphics[width=16.0cm]{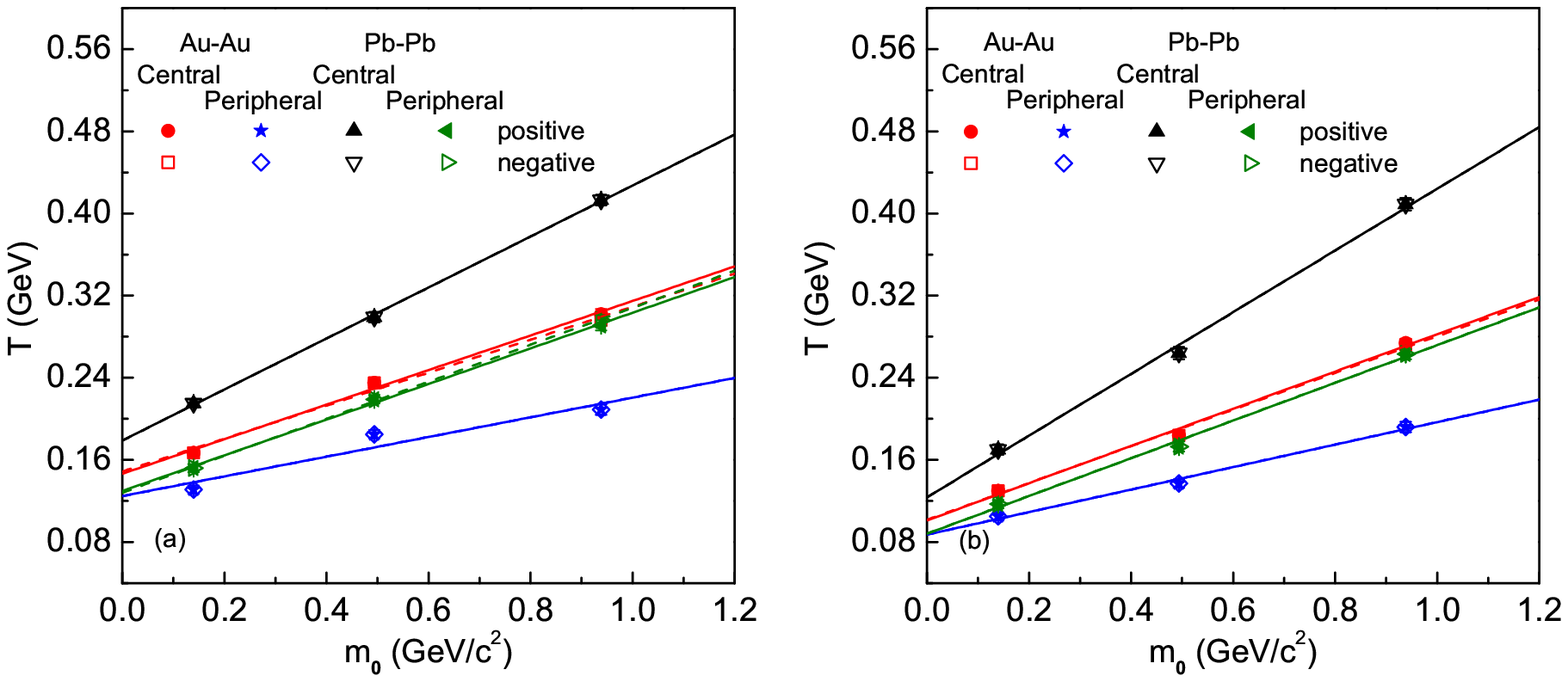}
\end{center}
\vskip0.5cm {\small Fig. 3. (Colour figure online) Relations
between $T$ and $m_0$. Different symbols represent central (0--5\%
and 0--12\%) and peripheral (80--92\% and 60--80\%) Au-Au
collisions at $\sqrt{s_{NN}}=200$ GeV and central (0--5\%) and
peripheral (80--90\% and 60--80\%) Pb-Pb collisions at
$\sqrt{s_{NN}}=2.76$ TeV respectively. The symbols presented in
panels (a) and (b) represent the results listed in Tables 4 and 5
and corresponded to the fits of Boltzmann and Tsallis
distributions respectively, where the closed and open symbols show
the results of positively and negatively charged particles
respectively. The solid and dashed lines are the results fitted by
the least square method for the positively and negatively charged
particles respectively, where the intercepts are regarded as
$T_0$.}
\end{figure*}

\begin{figure*}
\hskip-1.0cm \begin{center}
\includegraphics[width=16.0cm]{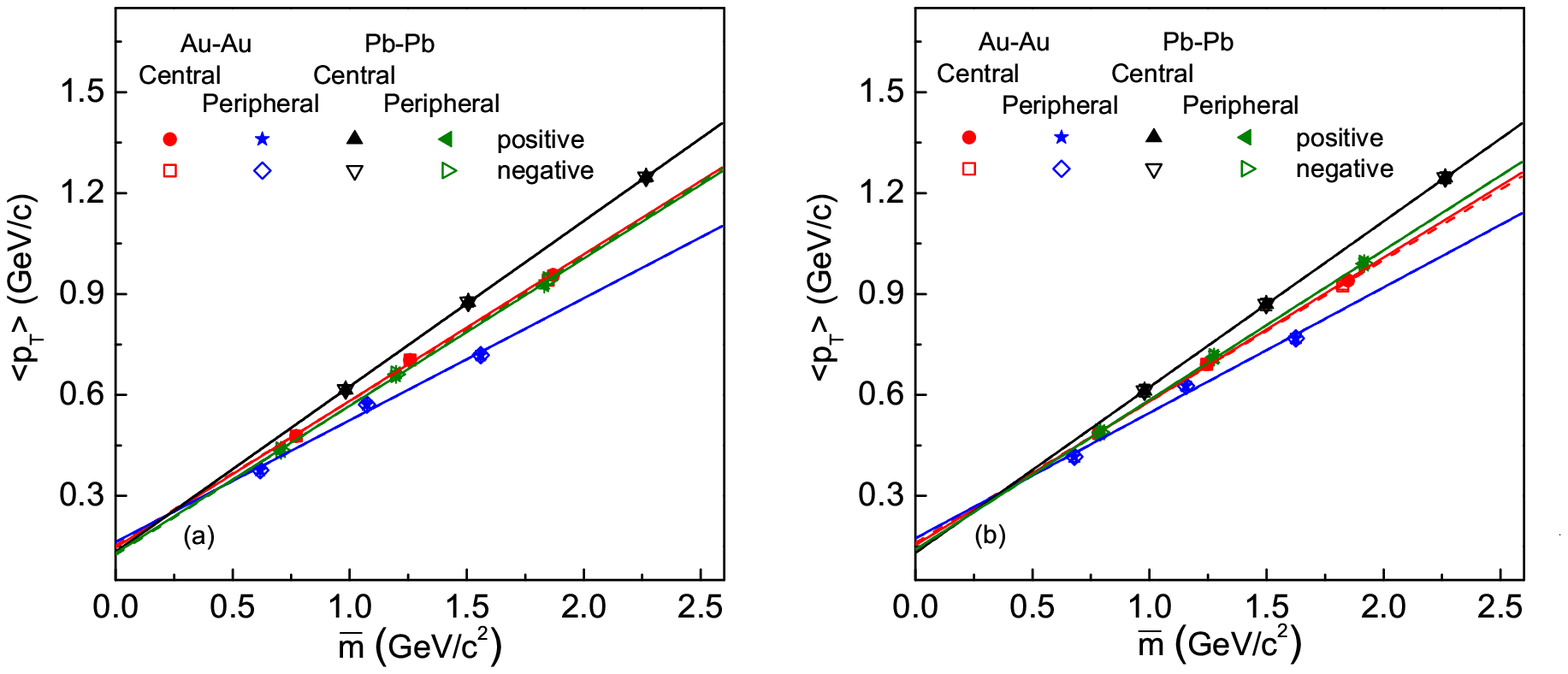}
\end{center}
\vskip0.5cm {\small Fig. 4. (Colour figure online) Same as Fig. 3,
but showing the relations between $\langle p_T \rangle$ and
$\overline{m}$, and the slopes are regarded as $\beta_T$. The
symbols presented in panels (a) and (b) represent the results
obtained according to the fits of Boltzmann and Tsallis
distributions respectively, where the values of parameters are
listed in Tables 4 and 5 respectively.}
\end{figure*}

\begin{figure*}
\hskip-1.0cm \begin{center}
\includegraphics[width=16.0cm]{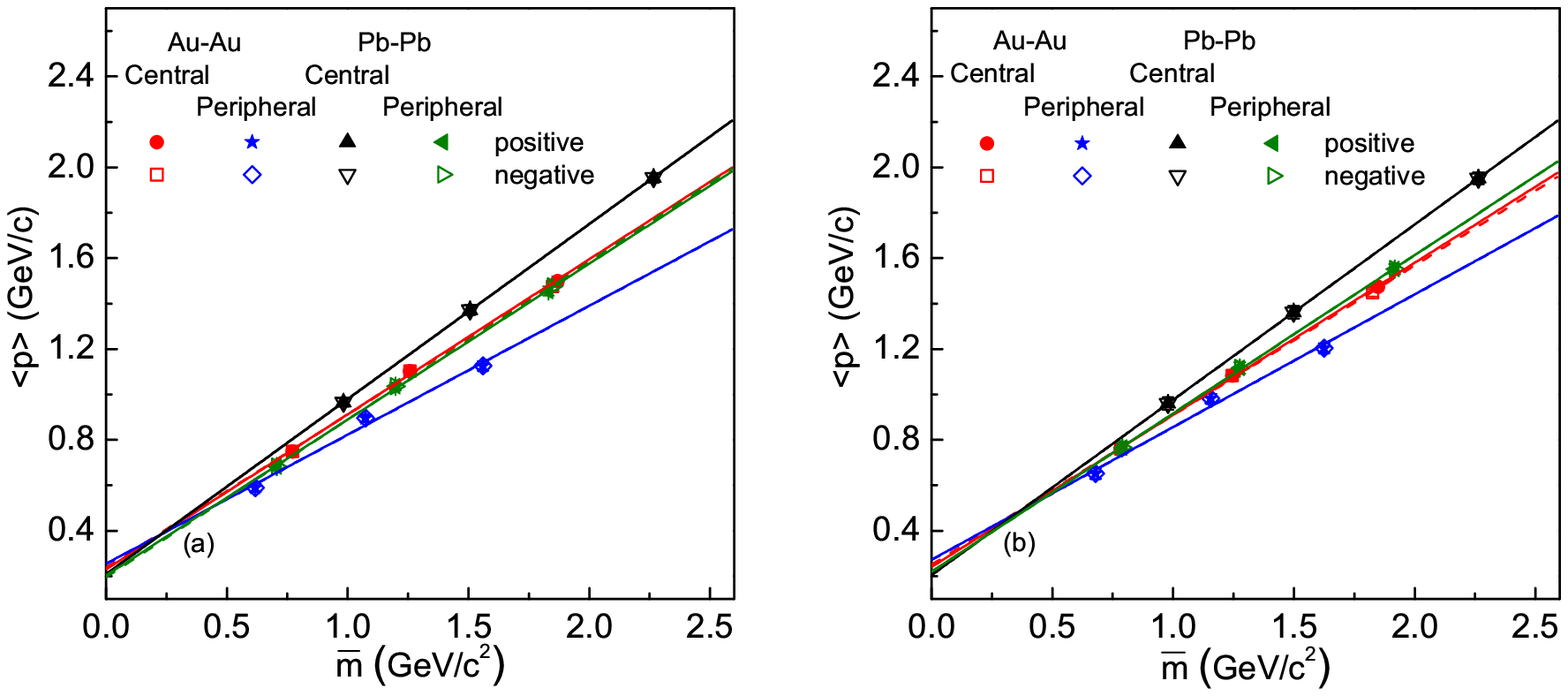}
\end{center}
\vskip0.5cm {\small Fig. 5. (Colour figure online) Same as Fig. 3,
but showing the relations between $\langle p \rangle$ and
$\overline{m}$, and the slopes are regarded as $\beta$. The
symbols presented in panels (a) and (b) represent the results
obtained according to the fits of Boltzmann and Tsallis
distributions respectively, where the values of parameters are
listed in Tables 4 and 5 respectively.}
\end{figure*}

\begin{table*}
{\scriptsize Table 6. Values of free parameters (intercept and
slope) and $\chi^2$/dof corresponding to the relations obtained
from the fits of the Boltzmann distribution in Figs. 3(a), 4(a),
and 5(a).
\begin{center}
\begin{tabular}{ccccccc}
\hline Figure & Relation & Type and main particles & Centrality & Intercept & Slope & $\chi^2$/dof \\
\hline
3(a) & $T-m_0$ & Au-Au positive &  Central  & $0.147\pm0.007$ & $0.168\pm0.012$ & $2.625$ \\
     &         &       negative &  Central  & $0.149\pm0.010$ & $0.160\pm0.016$ & $4.618$ \\
     &         &       positive & Peripheral& $0.125\pm0.017$ & $0.096\pm0.028$ & $14.910$\\
     &         &       negative & Peripheral& $0.125\pm0.017$ & $0.096\pm0.028$ & $14.910$\\
     &         & Pb-Pb positive &  Central  & $0.179\pm0.003$ & $0.248\pm0.005$ & $0.424$ \\
     &         &       negative &  Central  & $0.179\pm0.003$ & $0.248\pm0.005$ & $0.424$ \\
     &         &       positive & Peripheral& $0.130\pm0.005$ & $0.174\pm0.008$ & $1.142$ \\
     &         &       negative & Peripheral& $0.128\pm0.003$ & $0.180\pm0.005$ & $0.394$ \\
\hline
4(a) & $\langle p_T\rangle-\overline{m}$ & Au-Au positive &  Central  & $0.147\pm0.018$ & $0.436\pm0.013$ & $0.864$ \\
     &                                   &       negative &  Central  & $0.152\pm0.023$ & $0.430\pm0.017$ & $1.312$ \\
     &                                   &       positive & Peripheral& $0.163\pm0.041$ & $0.362\pm0.036$ & $4.734$ \\
     &                                   &       negative & Peripheral& $0.163\pm0.041$ & $0.362\pm0.036$ & $4.734$ \\
     &                                   & Pb-Pb positive &  Central  & $0.133\pm0.004$ & $0.492\pm0.002$ & $0.024$ \\
     &                                   &       negative &  Central  & $0.133\pm0.004$ & $0.492\pm0.002$ & $0.024$ \\
     &                                   &       positive & Peripheral& $0.130\pm0.013$ & $0.438\pm0.010$ & $0.499$ \\
     &                                   &       negative & Peripheral& $0.125\pm0.010$ & $0.443\pm0.007$ & $0.285$ \\
\hline
5(a) & $\langle p\rangle-\overline{m}$   & Au-Au positive &  Central  & $0.230\pm0.028$ & $0.683\pm0.021$ & $0.865$ \\
     &                                   &       negative &  Central  & $0.239\pm0.035$ & $0.673\pm0.026$ & $1.313$ \\
     &                                   &       positive & Peripheral& $0.255\pm0.064$ & $0.568\pm0.056$ & $4.746$ \\
     &                                   &       negative & Peripheral& $0.255\pm0.064$ & $0.568\pm0.056$ & $4.746$ \\
     &                                   & Pb-Pb positive &  Central  & $0.209\pm0.006$ & $0.771\pm0.003$ & $0.024$ \\
     &                                   &       negative &  Central  & $0.209\pm0.006$ & $0.771\pm0.003$ & $0.024$ \\
     &                                   &       positive & Peripheral& $0.203\pm0.020$ & $0.686\pm0.015$ & $0.496$ \\
     &                                   &       negative & Peripheral& $0.196\pm0.015$ & $0.694\pm0.011$ & $0.283$ \\
\hline
\end{tabular}
\end{center}}
\end{table*}

\begin{table*}
{\scriptsize Table 7. Values of free parameters (intercept and
slope) and $\chi^2$/dof corresponding to the relations obtained
from the fits of the Tsallis distribution in Figs. 3(b), 4(b), and
5(b).
\begin{center}
\begin{tabular}{ccccccc}
\hline Figure & Relation & Type and main particles & Centrality & Intercept & Slope & $\chi^2$/dof \\
\hline
3(b) & $T-m_0$ & Au-Au positive &  Central  & $0.101\pm0.009$ & $0.181\pm0.014$ & $3.059$ \\
     &         &       negative &  Central  & $0.102\pm0.008$ & $0.179\pm0.013$ & $2.533$ \\
     &         &       positive & Peripheral& $0.087\pm0.006$ & $0.110\pm0.009$ & $1.708$ \\
     &         &       negative & Peripheral& $0.087\pm0.006$ & $0.110\pm0.009$ & $1.708$ \\
     &         & Pb-Pb positive &  Central  & $0.124\pm0.011$ & $0.300\pm0.017$ & $2.877$ \\
     &         &       negative &  Central  & $0.124\pm0.011$ & $0.300\pm0.017$ & $2.877$ \\
     &         &       positive & Peripheral& $0.088\pm0.008$ & $0.184\pm0.013$ & $2.258$ \\
     &         &       negative & Peripheral& $0.088\pm0.008$ & $0.184\pm0.013$ & $2.258$ \\
\hline
4(b) & $\langle p_T\rangle-\overline{m}$ & Au-Au positive &  Central  & $0.154\pm0.013$ & $0.427\pm0.010$ & $0.270$ \\
     &                                   &       negative &  Central  & $0.160\pm0.018$ & $0.420\pm0.013$ & $0.495$ \\
     &                                   &       positive & Peripheral& $0.174\pm0.049$ & $0.373\pm0.040$ & $4.116$ \\
     &                                   &       negative & Peripheral& $0.174\pm0.049$ & $0.373\pm0.040$ & $4.116$ \\
     &                                   & Pb-Pb positive &  Central  & $0.131\pm0.001$ & $0.493\pm0.001$ & $0.001$ \\
     &                                   &       negative &  Central  & $0.131\pm0.001$ & $0.493\pm0.001$ & $0.001$ \\
     &                                   &       positive & Peripheral& $0.140\pm0.011$ & $0.445\pm0.008$ & $0.148$ \\
     &                                   &       negative & Peripheral& $0.140\pm0.011$ & $0.445\pm0.008$ & $0.148$ \\
\hline
5(b) & $\langle p\rangle-\overline{m}$   & Au-Au positive &  Central  & $0.240\pm0.021$ & $0.670\pm0.015$ & $0.269$ \\
     &                                   &       negative &  Central  & $0.251\pm0.028$ & $0.659\pm0.021$ & $0.494$ \\
     &                                   &       positive & Peripheral& $0.272\pm0.077$ & $0.584\pm0.063$ & $4.111$ \\
     &                                   &       negative & Peripheral& $0.272\pm0.077$ & $0.584\pm0.063$ & $4.111$ \\
     &                                   & Pb-Pb positive &  Central  & $0.205\pm0.002$ & $0.772\pm0.001$ & $0.001$ \\
     &                                   &       negative &  Central  & $0.205\pm0.002$ & $0.772\pm0.001$ & $0.001$ \\
     &                                   &       positive & Peripheral& $0.220\pm0.017$ & $0.697\pm0.012$ & $0.148$ \\
     &                                   &       negative & Peripheral& $0.220\pm0.017$ & $0.697\pm0.012$ & $0.148$ \\
\hline
\end{tabular}
\end{center}}
\end{table*}

\begin{figure*}
\hskip-1.0cm \begin{center}
\includegraphics[width=16.0cm]{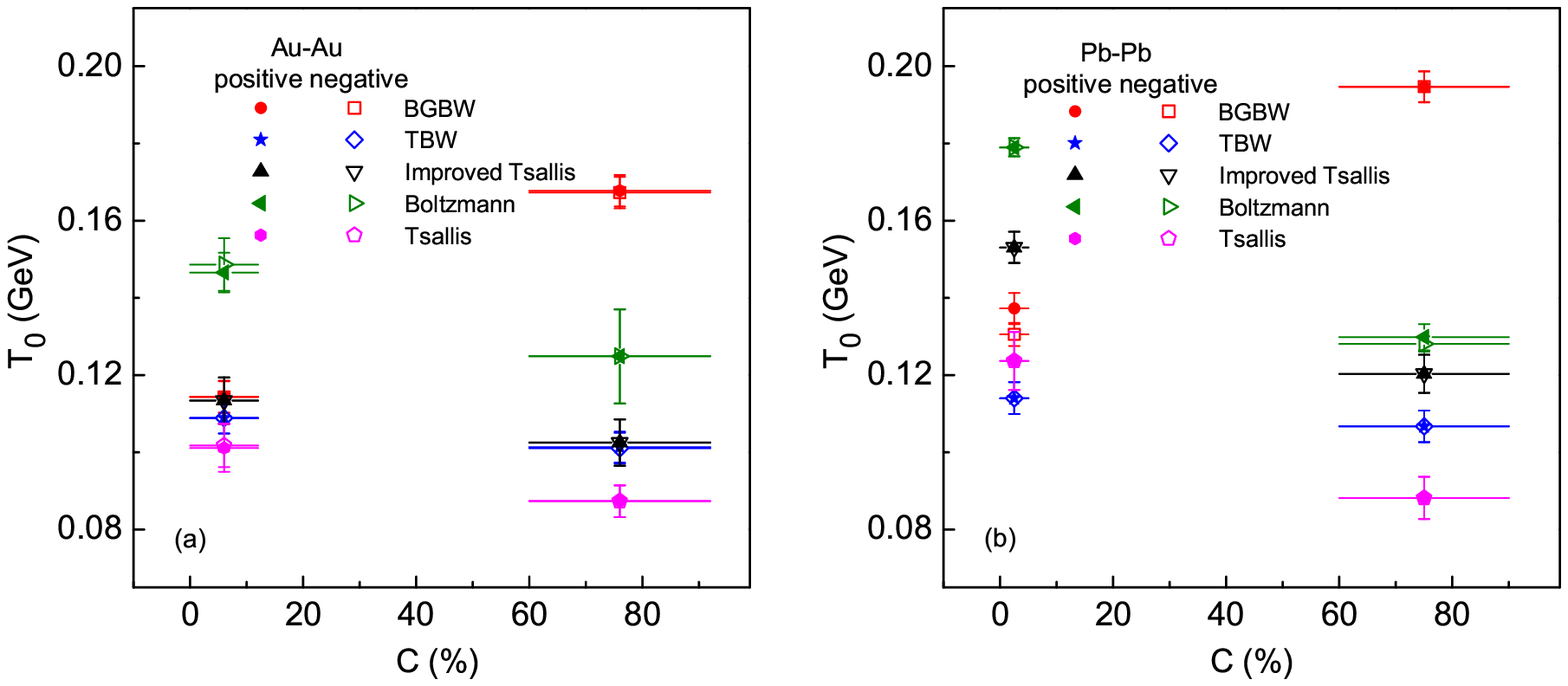}
\end{center}
\vskip0.5cm {\small Fig. 6. (Colour figure online) Comparisons of
$T_0$ obtained by different methods for different centralities
($C$), where the values of $T_0$ in the first three methods are
obtained by weighting different particles. Panels (a) and (b)
correspond to the results for central (0--5\% and 0--12\%) and
peripheral (80--92\% and 60--80\%) Au-Au collisions at
$\sqrt{s_{NN}}=200$ GeV and central (0--5\%) and peripheral
(80--90\% and 60--80\%) Pb-Pb collisions at $\sqrt{s_{NN}}=2.76$
TeV respectively, where the centralities 0--5\% and 0--12\%,
80--92\% and 60--80\%, as well as 80--90\% and 60--80\% are
combined to 0--12\%, 60--92\%, and 60--90\%, respectively.}
\end{figure*}

Figure 2 is the same as Fig. 1, but it shows the spectra,
$(1/N_{EV}) (2\pi p_T)^{-1} d^2N/(dydp_T)$, of (a)-(c) $\pi^+$
($\pi^++\pi^-$), $K^+$ ($K^++K^-$), and $p$ ($p+\bar p$), as well
as (b)-(d) $\pi^-$ ($\pi^++\pi^-$), $K^-$ ($K^++K^-$), $\bar p$
($p+\bar p$) produced in (a)-(b) central (0--5\%) and (c)-(d)
peripheral (80--90\% and 60--80\%) Pb-Pb collisions at
$\sqrt{s_{NN}}=2.76$ TeV, where $N_{EV}$ on the vertical axis
denotes the number of events, which is usually omitted. The closed
(open) symbols represent the experimental data of the ALICE
Collaboration measured in $|y|<0.5$ [28] (in $|\eta|<0.8$ for high
$p_T$ region and in $|y|<0.5$ for low $p_T$ region [29]). The data
for $\pi^++\pi^-$, $K^++K^-$, and $p+\bar p$ in (a)-(c) and
(b)-(d) are the same. One can see that, in most cases, all of the
considered methods describe approximately the $p_T$ spectra of
identified particles produced in central and peripheral Pb-Pb
collisions at $\sqrt{s_{NN}}=2.76$ TeV. Because the values of
$\chi^2$/dof in most cases are greater than 2 and sometimes as
large as 20.5, the fits in Figs. 1 and 2 are only approximate and
qualitative. The large values of $\chi^2$/dof in the present work
are caused by two factors which are the very small errors in the
data and large dispersion between the curve and data in some
cases. It is hard to reduce the values of $\chi^2$/dof in our
fits.

In the above fits, we have an addition term of inverse power-law
to account for hard process. This part contributes a small
fraction to the $p_T$ spectra, though the contribution coverage is
wide. In the fitting procedure, according to the changing tendency
of data in a low $p_T$ range from 0 to 2 GeV/$c$, the part for
soft process can be well constrained first of all, though the
contribution of soft process can even reach to 3.5 GeV/$c$. Then,
the part for hard process can be also constrained conveniently. In
addition, in order to give a set of fitted parameters as
accurately as possible, we use the least square method in the
whole $p_T$ coverage. It seems that different fitted parameters
can be obtained in different $p_T$ coverages. We should use a
$p_T$ coverage as widely as possible, especially for the
extraction of the parameters related to the inverse power-law
because that a limited $p_T$ coverage can not provide a good
constrain of the inverse power-law and thus can easily drive the
fitted parameters away from their physical meanings. In fact, for
extractions of the effective temperature and transverse flow
velocity which are the main topics of the present work, a not too
wide $p_T$ coverage such as 0--2$\sim$3 GeV/$c$ is enough due to
the soft process contributing only in the low $p_T$ region and the
changing tendency of data in 0--2 GeV/$c$ takes part in a main
role.

From the above fits one can see that, as a two-component function,
Eq. (9) with different soft components can approximately describe
the data in a wide $p_T$ coverage. In addition, in our very recent
work [37], we used the method iii) to describe preliminarily the
$p_T$ spectra up to nearly 20 GeV/$c$. In our another work [38],
two-Boltzmann distribution was used to describe the $p_T$ spectra
up to nearly 14 GeV/$c$. Generally, different sets of parameters
are needed for different data. In particular, as it is pointed out
in Ref. [39], more fitting parameters are needed in order to fit
wider $p_T$ range of particle spectra. In the present work, we fit
the particle spectra in a wide $p_T$ range by introducing the
inverse power-law to describe the high $p_T$ region. The price to
pay is 3 more parameters are added. In the two-component function,
the contributions of soft and hard components have a little effect
in constraining respective free parameters due to different
contributive regions, though the contribution fraction of the two
components is main role. This results in the $p_T$ coverage having
a small effect on $T_0$ and $\beta_T$. In fact, if we change the
boundary of low $p_T$ region from 2 to 3 or 3.5 GeV/$c$, the
variations of parameters can be neglected due to the tendency of
curve being mainly determined by the data in 0--2 GeV/$c$.
Meanwhile, the data in 2--3.5 GeV/$c$ obey naturally the tendency
of curve due to also the contribution or revision of hard
component. In other words, because of the revision of hard
component, the values of $T_0$ and $\beta_T$ are not sensitive to
the boundary of low $p_T$ region. Although different $p_T$
coverages obtained in different conditions can drive different
fitted curves, these differences appear mainly in the high $p_T$
region and do not largely effect the extraction of $T_0$ and
$\beta_T$. In any case, we always use the last square method to
extract the fitted parameters. In fact, the method used by us has
the minimum randomness in the extractions of the fitted
parameters.

It should be noted that although the conventional BGBW and TBW
models have only 2--3 parameters to describe the $p_T$ shape and
usually fit several spectra simultaneously to reduce the
correlation of the parameters, they seems to cover
non-simultaneity of the kinetic freeze-outs of different
particles. In the present work, although we use 3 more parameters
and fit each spectrum individually, we observe an evidence of the
mass dependent differential kinetic freeze-out scenario or
multiple kinetic freeze-outs scenario [4, 16, 23]. The larger the
temperature (mass) is, the earlier the particle produces. The
average temperature (flow velocity and entropy index) of the
kinetic freeze-outs for different particles is obtained by
weighting different $T_0$ ($\beta_T$ and $q$), where the weight
factor is the normalization constant of each $p_T$ spectrum. In
the case of using the average temperature (flow velocity and
entropy index) to fit the pion, kaon, and proton simultaneously to
better constrain the parameters, larger values of $\chi^2$/dof are
obtained.

Based on the descriptions of $p_T$ spectra, the first three
methods can give $T_0$ and $\beta_T$ conveniently, though the
values of parameters are possibly inharmonious due to different
methods. In particular, the value of $T_0$ obtained by the method
i) in peripheral collisions is larger than that in central
collisions, which is different from the methods ii) and iii) which
obtain an opposite result. According to the conventional treatment
in refs. [11, 14], the values of $\beta_T$ obtained by the methods
i) and ii) in peripheral collisions are taken to be nearly zero,
which are different from the method iii) which obtains a value of
about 0.6$c$ in both central and peripheral collisions.

To obtain the values of $T_0$, $\beta_T$, and $\beta$ by the
methods iv)$_{\rm a}$ and iv)$_{\rm b}$, we analyze the values of
$T$ presented in Tables 4 and 5, and calculate $\langle p_T
\rangle$, $\langle p \rangle$, and $\overline{m}$ based on the
values of parameters listed in Tables 4 and 5. In the calculations
performed from $p_T$ to $\langle p \rangle$ and $\overline{m}$ by
the Monte Carlo method, an isotropic assumption in the rest frame
of emission source is used [22--24]. In particular, $\overline{m}$
is in fact the mean energy $\langle \sqrt{p^2+m_0^2} \rangle$.

The relations between $T$ and $m_0$, $\langle p_T \rangle$ and
$\overline{m}$, as well as $\langle p \rangle$ and $\overline{m}$
are shown in Figs. 3, 4, and 5, respectively, where panels (a) and
(b) correspond to the methods iv)$_{\rm a}$ and iv)$_{\rm b}$
which use the Boltzmann and Tsallis distributions respectively.
Different symbols represent central (0--5\% and 0--12\%) and
peripheral (80--92\% and 60--80\%) Au-Au collisions at
$\sqrt{s_{NN}}=200$ GeV and central (0--5\%) and peripheral
(80--90\% and 60--80\%) Pb-Pb collisions at $\sqrt{s_{NN}}=2.76$
TeV respectively, where the centralities 0--5\% and 0--12\%,
80--92\% and 60--80\%, as well as 80--90\% and 60--80\% can be
combined to 0--12\%, 60--92\%, and 60--90\%, respectively. The
symbols in Fig. 3 represent values of $T$ listed in Tables 4 and 5
for different $m_0$. The symbols in Figs. 4 and 5 represent values
of $\langle p_T \rangle$ and $\langle p \rangle$ for different
$\overline{m}$ respectively, which are calculated due to the
parameters listed in Tables 4 and 5 and the isotropic assumption
in the rest frame of emission source. The solid and dashed lines
in the three figures are the results fitted by the least square
method for the positively and negatively charged particles
respectively. The values of intercepts, slopes, and $\chi^2$/dof
are listed in Tables 6 and 7 which correspond to the methods
iv)$_{\rm a}$ and iv)$_{\rm b}$ respectively. One can see that, in
most cases, the mentioned relations are described by a linear
function. In particular, the intercept in Fig. 3 is regarded as
$T_0$, and the slopes in Figs. 4 and 5 are regarded as $\beta_T$
and $\beta$ respectively. The values of $T$, $T_0$, $\beta_T$,
$\beta$, and $\overline{m}$ are approximately independent of
isospin.

\begin{figure*}
\hskip-1.0cm \begin{center}
\includegraphics[width=16.0cm]{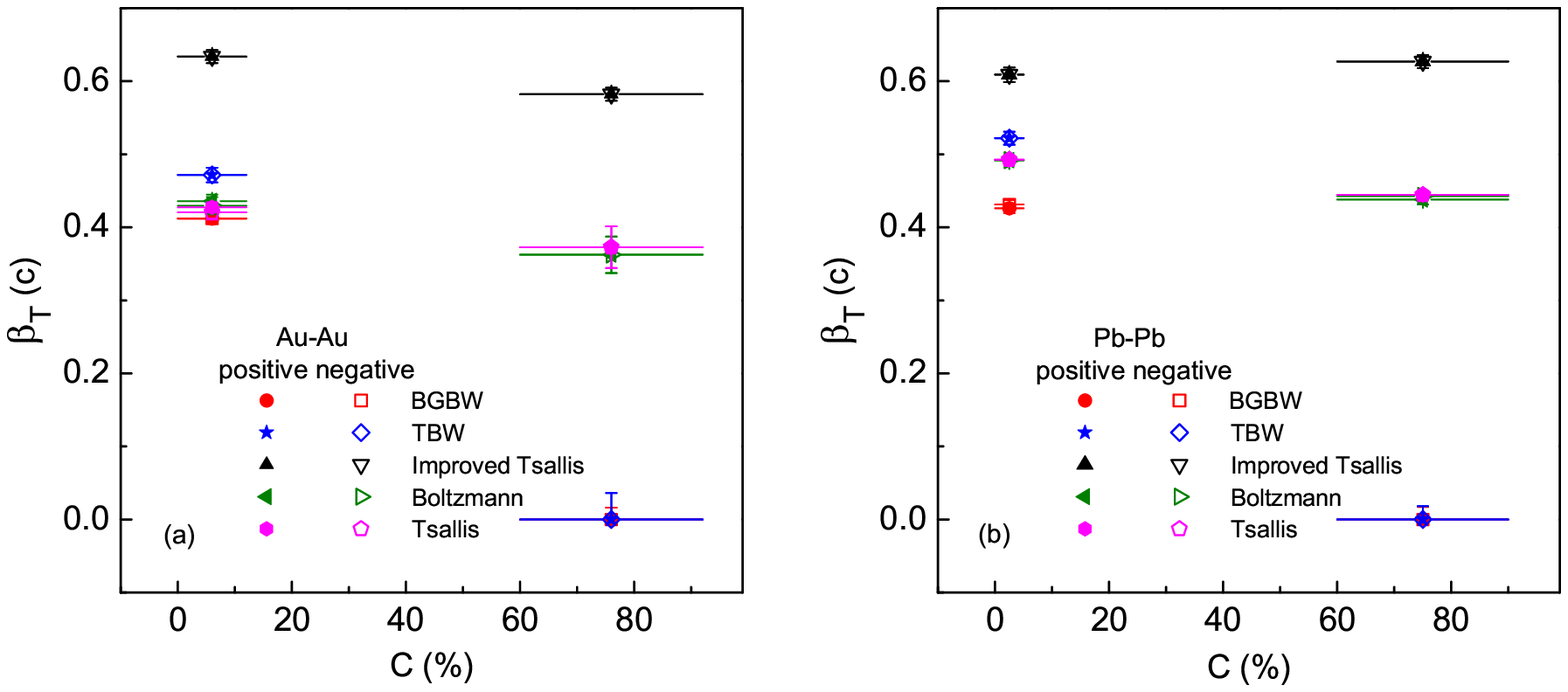}
\end{center}
\vskip0.5cm {\small Fig. 7. (Colour figure online) Same as Fig. 6,
but showing the comparisons of $\beta_T$ obtained by different
methods for different centralities.}
\end{figure*}

To compare values of key parameters obtained by different methods
for different centralities (both central and peripheral
collisions), Figs. 6 and 7 show $T_0$ and $\beta_T$ respectively,
where panels (a) and (b) correspond to the results for central
(0--5\% and 0--12\%) and peripheral (80--92\% and 60--80\%) Au-Au
collisions at $\sqrt{s_{NN}}=200$ GeV and central (0--5\%) and
peripheral (80--90\% and 60--80\%) Pb-Pb collisions at
$\sqrt{s_{NN}}=2.76$ TeV respectively. The closed and open symbols
represent positively and negatively charged particles
respectively, which are quoted from Tables 1, 2, 3, 6, and 7 which
correspond to the methods i), ii), iii), iv)$_{\rm a}$, and
iv)$_{\rm b}$, respectively. In particular, the values of $T_0$
and $\beta_T$ in the first three methods are obtained by weighting
different particles. One can see that, by using the method i), the
value of $T_0$ in central collisions is smaller than that in
peripheral collisions, and other methods present a larger $T_0$ in
central collisions. The methods i) and ii) show a nearly zero
$\beta_T$ in peripheral collisions according to refs. [11, 14],
while other methods show a considerable $\beta_T$ in both central
and peripheral collisions.

To explain the inconsistent results in $T_0$ and $\beta_T$ for
different methods, we re-examine the first two methods. It should
be noticed that the same flow profile function,
$\beta(r)=\beta_S(r/R)^{n_0}$, and the same transverse flow
velocity, $\beta_T=2\beta_S/(n_0+2)$, are used in the first two
methods, though $n_0=2$ is used in the method i) [11] and $n_0=1$
is used in the method ii) [14] with the conventional treatment. As
an insensitive quantity, although the radial size $R$ of the
thermal source in central collisions can be approximately regarded
as the radius of a collision nucleus, and in peripheral collisions
$R$ is not zero due to a few participant nucleons taking part in
the interactions in which we can take approximately $R$ to be 2.5
fm, both the methods i) and ii) use a nearly zero $\beta_T$ in
peripheral collisions [11, 14]. If we consider a non-zero
$\beta_T$ in peripheral collisions for the methods i) and ii), the
situation will be changed.

By using a non-zero $\beta_T$ in peripheral collisions for the
methods i) and ii), we re-analyze the data presented in Figs. 1
and 2. At the same time, to see the influences of different $n_0$
in the self-similar flow profile, we refit the mentioned $p_T$
spectra by the first two methods with $n_0=1$ and 2 synchronously.
The results re-analyzed by us are shown in Figs. 8 and 9 which
correspond to 200 GeV Au-Au and 2.76 TeV Pb-Pb collisions
respectively. The data points are the same as Figs. 1 and 2
[25--29]. The dotted, solid, dashed, and dotted-dashed curves
correspond to the results of the method i) with $n_0=1$ and 2, and
of the method ii) with $n_0=1$ and 2, respectively, where the
results of the method i) with $n_0=2$ and of the method ii) with
$n_0=1$ in central collisions are the same as Figs. 1 and 2. The
values of related parameters and $\chi^2$/dof are listed in Tables
8 and 9, where the parameters for the method i) with $n_0=2$ and
for the method ii) with $n_0=1$ in central collisions repeat those
in Tables 1 and 2, which are not listed again. One can see that,
after the re-examination, the values of $T_0$ in central
collisions are larger than those in peripheral collisions. The
values of $\beta_T$ in peripheral collisions are no longer zero.
These new results are consistent with other methods.

\begin{figure*}
\hskip-1.0cm
\begin{center}
\includegraphics[width=16.0cm]{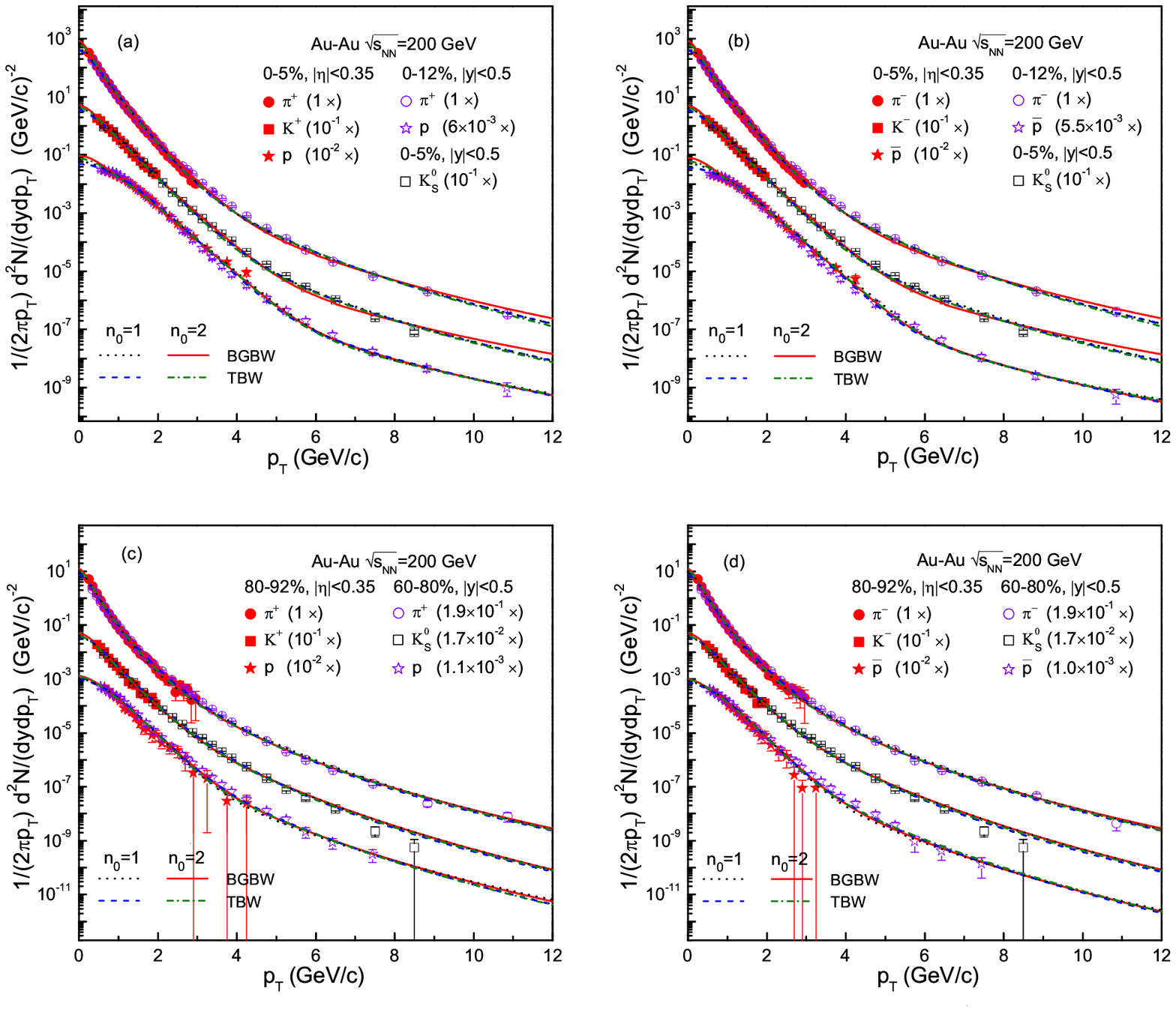}
\end{center}
\vskip0.5cm {\small Fig. 8. (Colour figure online) Reanalyzing the
transverse momentum spectra [25--27] collected in Fig. 1 by the
first two methods. The dotted, solid, dashed, and dotted-dashed
curves are our results calculated by using the method i) with
$n_0=1$ and 2, as well as the method ii) with $n_0=1$ and 2,
respectively. The results for central collisions obtained by the
method i) with $n_0=2$ and by the method ii) with $n_0=1$ are the
same as Fig. 1.}
\end{figure*}

\begin{figure*}
\hskip-1.0cm
\begin{center}
\includegraphics[width=16.0cm]{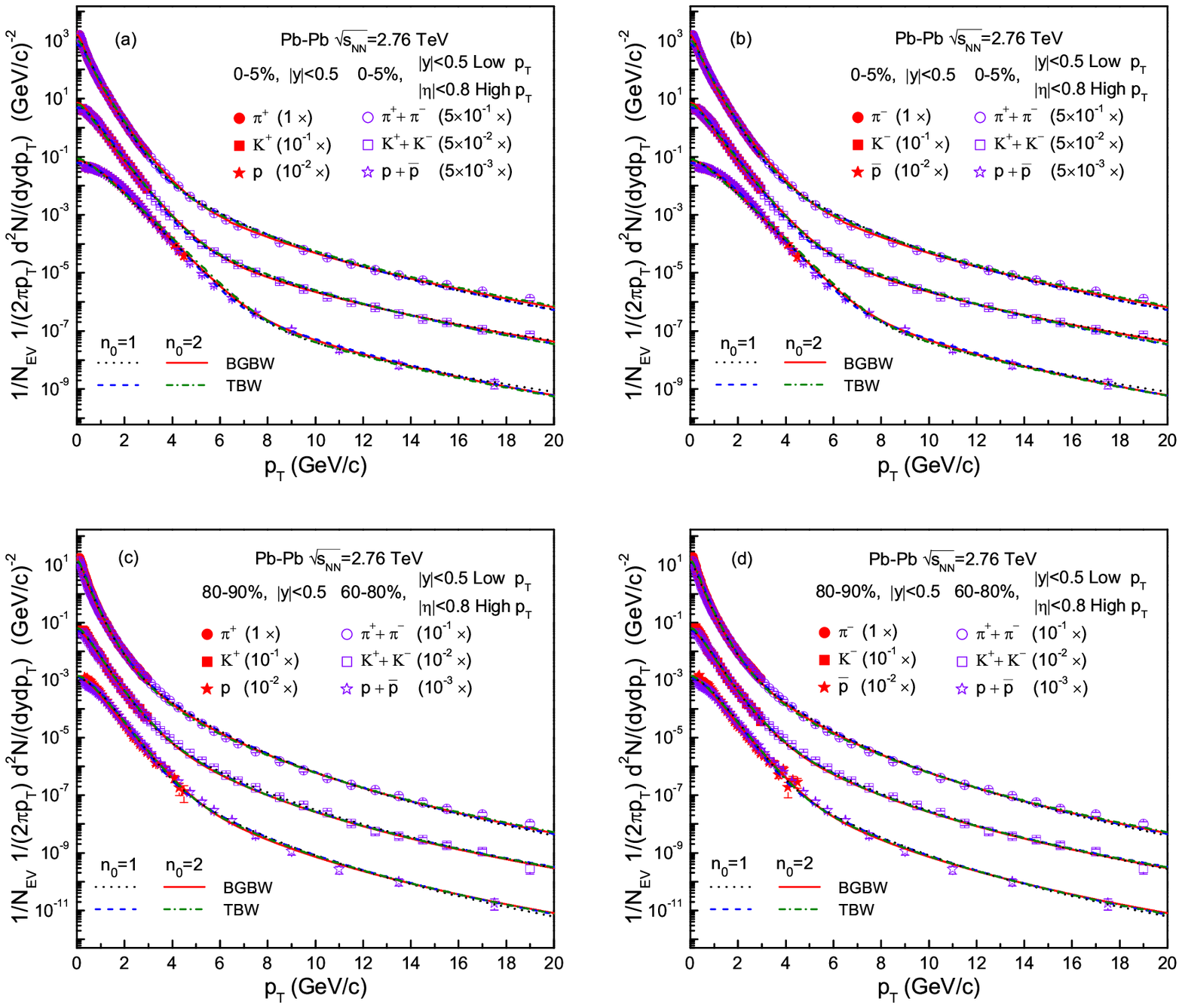}
\end{center}
\vskip0.5cm {\small Fig. 9. (Colour figure online) Same as Fig. 8,
but reanalyzing the transverse momentum spectra [28, 29] collected
in Fig. 2 by the first two methods. The results for central
collisions obtained by the method i) with $n_0=2$ and by the
method ii) with $n_0=1$ are the same as Fig. 2.}
\end{figure*}

\begin{table*}
{\tiny Table 8. Values of free parameters ($T_0$, $\beta_{T}$,
$k$, $p_{0}$, and $n$), normalization constant ($N_{0}$), and
$\chi^2$/dof corresponding to the fits of method i) in Figs. 8 and
9, where the values for central collisions with $n_0=2$ in the
self-similar flow profile repeat those in Table 1, which are not
listed again.
\begin{center}
\begin{tabular}{cccccccccc}
\hline Fig. & Cent. & Main Part. & $T_0$ (GeV) & $\beta_{T}$ ($c$) & $k$ & $p_{0}$ (GeV/$c$) & $n$ & $N_0$ & $\chi^2$/dof \\
\hline
8(a)    &  Central  & $\pi^+$   & $0.138\pm0.005$ & $0.452\pm0.008$ & $0.964\pm0.006$ & $2.375\pm0.069$ & $10.365\pm0.188$ & $633.869\pm62.976$ & $3.369$\\
Au-Au   &           & $K^+$     & $0.169\pm0.005$ & $0.412\pm0.008$ & $0.901\pm0.006$ & $1.998\pm0.058$ & $9.675\pm0.185$  & $54.966\pm3.838$   & $5.502$\\
$n_0=1$ &           & $p$       & $0.198\pm0.005$ & $0.398\pm0.008$ & $0.995\pm0.002$ & $2.485\pm0.072$ & $8.075\pm0.171$  & $8.457\pm0.646$    & $5.274$\\
8(b)    &  Central  & $\pi^-$   & $0.138\pm0.005$ & $0.452\pm0.008$ & $0.964\pm0.006$ & $2.375\pm0.069$ & $10.365\pm0.188$ & $633.869\pm62.976$ & $3.277$\\
        &           & $K^-$     & $0.169\pm0.005$ & $0.412\pm0.008$ & $0.901\pm0.006$ & $2.098\pm0.060$ & $9.835\pm0.188$  & $54.759\pm3.823$   & $6.405$\\
        &           & $\bar{p}$ & $0.198\pm0.005$ & $0.397\pm0.008$ & $0.994\pm0.002$ & $2.185\pm0.070$ & $7.975\pm0.168$  & $7.096\pm0.649$    & $12.058$\\
8(c)    & Peripheral& $\pi^+$   & $0.115\pm0.005$ & $0.415\pm0.008$ & $0.901\pm0.008$ & $2.512\pm0.079$ & $11.123\pm0.173$ & $11.713\pm0.591$   & $4.455$\\
        &           & $K^+$     & $0.145\pm0.005$ & $0.415\pm0.008$ & $0.888\pm0.008$ & $3.923\pm0.082$ & $12.923\pm0.178$ & $0.482\pm0.077$    & $6.711$\\
        &           & $p$       & $0.157\pm0.006$ & $0.353\pm0.008$ & $0.947\pm0.008$ & $3.316\pm0.069$ & $11.016\pm0.169$ & $0.142\pm0.015$    & $1.444$\\
8(d)    & Peripheral& $\pi^-$   & $0.115\pm0.005$ & $0.415\pm0.008$ & $0.901\pm0.008$ & $2.512\pm0.079$ & $11.123\pm0.173$ & $11.713\pm0.591$   & $3.800$\\
        &           & $K^-$     & $0.145\pm0.005$ & $0.415\pm0.008$ & $0.888\pm0.008$ & $3.923\pm0.082$ & $12.923\pm0.178$ & $0.482\pm0.077$    & $5.907$\\
        &           & $\bar{p}$ & $0.157\pm0.006$ & $0.353\pm0.008$ & $0.945\pm0.008$ & $3.316\pm0.069$ & $11.528\pm0.169$ & $0.112\pm0.011$    & $0.904$\\
\hline
8(c)    & Peripheral& $\pi^+$   & $0.103\pm0.005$ & $0.395\pm0.008$ & $0.896\pm0.008$ & $2.012\pm0.063$ & $10.203\pm0.185$ & $14.240\pm1.308$   & $2.956$\\
Au-Au   &           & $K^+$     & $0.117\pm0.006$ & $0.383\pm0.008$ & $0.901\pm0.008$ & $3.983\pm0.071$ & $12.993\pm0.195$ & $0.636\pm0.033$    & $4.221$\\
$n_0=2$ &           & $p$       & $0.118\pm0.006$ & $0.355\pm0.008$ & $0.905\pm0.008$ & $3.268\pm0.066$ & $11.506\pm0.186$ & $0.170\pm0.012$    & $1.093$\\
8(d)    & Peripheral& $\pi^-$   & $0.103\pm0.005$ & $0.395\pm0.008$ & $0.896\pm0.008$ & $2.012\pm0.063$ & $10.203\pm0.185$ & $14.240\pm1.308$   & $2.652$\\
        &           & $K^-$     & $0.117\pm0.006$ & $0.383\pm0.008$ & $0.901\pm0.008$ & $3.983\pm0.071$ & $12.993\pm0.195$ & $0.636\pm0.033$    & $3.879$\\
        &           & $\bar{p}$ & $0.118\pm0.006$ & $0.355\pm0.008$ & $0.905\pm0.008$ & $3.268\pm0.066$ & $11.926\pm0.186$ & $0.128\pm0.012$    & $0.589$\\
\hline
9(a)    &  Central  & $\pi^+$   & $0.149\pm0.005$ & $0.473\pm0.008$ & $0.922\pm0.008$ & $1.535\pm0.056$ & $7.276\pm0.104$ & $1465.409\pm127.197$  & $3.815$\\
Pb-Pb   &           & $K^+$     & $0.235\pm0.005$ & $0.399\pm0.008$ & $0.938\pm0.008$ & $1.295\pm0.055$ & $6.114\pm0.101$ & $77.086\pm7.666$      & $1.463$\\
$n_0=1$ &           & $p$       & $0.338\pm0.005$ & $0.332\pm0.006$ & $0.991\pm0.002$ & $2.285\pm0.082$ & $6.485\pm0.108$ & $10.152\pm0.330$      & $11.411$\\
9(b)    &  Central  & $\pi^-$   & $0.149\pm0.005$ & $0.473\pm0.008$ & $0.922\pm0.008$ & $1.535\pm0.056$ & $7.276\pm0.104$ & $1465.409\pm127.197$  & $3.751$\\
        &           & $K^-$     & $0.235\pm0.005$ & $0.399\pm0.008$ & $0.938\pm0.008$ & $1.295\pm0.055$ & $6.114\pm0.101$ & $77.157\pm7.674$      & $1.229$\\
        &           & $\bar{p}$ & $0.338\pm0.005$ & $0.332\pm0.006$ & $0.991\pm0.002$ & $2.285\pm0.082$ & $6.485\pm0.108$ & $10.152\pm0.330$      & $10.234$\\
9(c)    & Peripheral& $\pi^+$   & $0.127\pm0.005$ & $0.473\pm0.008$ & $0.934\pm0.008$ & $2.793\pm0.078$ & $8.765\pm0.138$ & $14.233\pm0.756$      & $8.290$\\
        &           & $K^+$     & $0.169\pm0.004$ & $0.453\pm0.008$ & $0.902\pm0.008$ & $2.665\pm0.074$ & $7.995\pm0.129$ & $0.723\pm0.050$       & $2.448$\\
        &           & $p$       & $0.180\pm0.005$ & $0.436\pm0.008$ & $0.918\pm0.008$ & $2.995\pm0.092$ & $8.599\pm0.132$ & $0.167\pm0.014$       & $3.944$\\
9(d)    & Peripheral& $\pi^-$   & $0.127\pm0.005$ & $0.473\pm0.008$ & $0.934\pm0.008$ & $2.793\pm0.078$ & $8.765\pm0.138$ & $14.233\pm0.756$      & $8.285$\\
        &           & $K^-$     & $0.169\pm0.004$ & $0.453\pm0.008$ & $0.902\pm0.008$ & $2.665\pm0.074$ & $7.995\pm0.129$ & $0.723\pm0.050$       & $2.686$\\
        &           & $\bar{p}$ & $0.180\pm0.005$ & $0.436\pm0.008$ & $0.918\pm0.008$ & $2.995\pm0.092$ & $8.599\pm0.132$ & $0.167\pm0.014$       & $4.196$\\
\hline
9(c)    & Peripheral& $\pi^+$   & $0.116\pm0.004$ & $0.410\pm0.008$ & $0.941\pm0.007$ & $2.393\pm0.058$ & $8.185\pm0.153$ & $17.976\pm0.731$      & $4.533$\\
Pb-Pb   &           & $K^+$     & $0.184\pm0.005$ & $0.367\pm0.008$ & $0.908\pm0.007$ & $2.375\pm0.056$ & $7.585\pm0.145$ & $0.702\pm0.044$       & $1.120$\\
$n_0=2$ &           & $p$       & $0.204\pm0.005$ & $0.343\pm0.008$ & $0.919\pm0.007$ & $2.178\pm0.055$ & $7.515\pm0.145$ & $0.172\pm0.015$       & $1.791$\\
9(d)    & Peripheral& $\pi^-$   & $0.116\pm0.004$ & $0.410\pm0.008$ & $0.941\pm0.007$ & $2.393\pm0.058$ & $8.185\pm0.153$ & $17.976\pm0.731$      & $4.601$\\
        &           & $K^-$     & $0.184\pm0.005$ & $0.367\pm0.008$ & $0.908\pm0.007$ & $2.375\pm0.056$ & $7.585\pm0.145$ & $0.702\pm0.044$       & $1.232$\\
        &           & $\bar{p}$ & $0.204\pm0.005$ & $0.343\pm0.008$ & $0.919\pm0.007$ & $2.178\pm0.055$ & $7.515\pm0.145$ & $0.172\pm0.015$       & $1.963$\\
\hline
\end{tabular}
\end{center}}
\end{table*}

\begin{table*}
{\tiny Table 9. Values of free parameters ($T_0$, $q$,
$\beta_{T}$, $k$, $p_{0}$, and $n$), normalization constant
($N_{0}$), and $\chi^2$/dof corresponding to the fits of method
ii) in Figs. 8 and 9, where the values for central collisions with
$n_0=1$ in the self-similar flow profile repeat those in Table 2,
which are not listed again.
\begin{center}
\begin{tabular}{ccccccccccc}
\hline Fig. & Cent. & Main Part. & $T_0$ (GeV) & $q$ & $\beta_{T}$ ($c$) & $k$ & $p_{0}$ (GeV/$c$) & $n$ & $N_0$ & $\chi^2$/dof \\
\hline
8(c)    & Peripheral& $\pi^+$   & $0.079\pm0.004$ & $1.069\pm0.006$ & $0.405\pm0.009$ & $0.924\pm0.006$ & $2.192\pm0.083$ & $10.379\pm0.189$ & $9.197\pm0.912$    & $1.715$ \\
Au-Au   &           & $K^+$     & $0.089\pm0.005$ & $1.063\pm0.005$ & $0.389\pm0.009$ & $0.921\pm0.006$ & $3.602\pm0.096$ & $12.282\pm0.165$ & $0.491\pm0.052$    & $4.499$ \\
$n_0=1$ &           & $p$       & $0.095\pm0.005$ & $1.028\pm0.005$ & $0.389\pm0.009$ & $0.902\pm0.007$ & $3.810\pm0.102$ & $12.568\pm0.171$ & $0.134\pm0.010$    & $1.457$ \\
8(d)    & Peripheral& $\pi^-$   & $0.079\pm0.004$ & $1.069\pm0.006$ & $0.405\pm0.009$ & $0.924\pm0.006$ & $2.192\pm0.083$ & $10.379\pm0.189$ & $9.197\pm0.912$    & $1.445$ \\
        &           & $K^-$     & $0.089\pm0.005$ & $1.061\pm0.005$ & $0.389\pm0.009$ & $0.921\pm0.006$ & $3.602\pm0.096$ & $12.282\pm0.165$ & $0.486\pm0.052$    & $3.127$ \\
        &           & $\bar{p}$ & $0.095\pm0.005$ & $1.028\pm0.005$ & $0.389\pm0.009$ & $0.908\pm0.007$ & $3.810\pm0.102$ & $12.868\pm0.171$ & $0.100\pm0.010$    & $0.670$ \\
\hline
8(a)    &  Central  & $\pi^+$   & $0.091\pm0.003$ & $1.010\pm0.005$ & $0.401\pm0.008$ & $0.985\pm0.003$ & $3.591\pm0.091$ & $12.035\pm0.173$ & $683.617\pm48.090$ & $3.630$ \\
Au-Au   &           & $K^+$     & $0.103\pm0.005$ & $1.008\pm0.004$ & $0.395\pm0.007$ & $0.961\pm0.004$ & $2.675\pm0.103$ & $10.327\pm0.089$ & $51.119\pm5.034$   & $5.703$ \\
$n_0=2$ &           & $p$       & $0.118\pm0.005$ & $1.009\pm0.004$ & $0.374\pm0.005$ & $0.997\pm0.002$ & $3.385\pm0.168$ & $8.895\pm0.108$  & $9.706\pm0.421$    & $6.866$\\
8(b)    &  Central  & $\pi^-$   & $0.091\pm0.003$ & $1.010\pm0.005$ & $0.401\pm0.008$ & $0.985\pm0.003$ & $3.591\pm0.091$ & $12.035\pm0.173$ & $683.617\pm48.090$ & $3.362$ \\
        &           & $K^-$     & $0.103\pm0.005$ & $1.008\pm0.004$ & $0.395\pm0.007$ & $0.961\pm0.004$ & $2.675\pm0.103$ & $10.327\pm0.159$ & $49.059\pm5.034$   & $6.731$ \\
        &           & $\bar{p}$ & $0.118\pm0.005$ & $1.009\pm0.004$ & $0.374\pm0.005$ & $0.997\pm0.002$ & $3.385\pm0.168$ & $9.095\pm0.112$  & $7.862\pm0.422$    & $15.669$\\
8(c)    & Peripheral& $\pi^+$   & $0.073\pm0.004$ & $1.025\pm0.004$ & $0.398\pm0.008$ & $0.943\pm0.004$ & $2.653\pm0.091$ & $11.093\pm0.169$ & $10.627\pm0.888$   & $3.602$ \\
        &           & $K^+$     & $0.082\pm0.005$ & $1.033\pm0.005$ & $0.380\pm0.008$ & $0.891\pm0.005$ & $3.683\pm0.092$ & $12.553\pm0.170$ & $0.470\pm0.005$    & $4.498$ \\
        &           & $p$       & $0.085\pm0.005$ & $1.009\pm0.005$ & $0.359\pm0.008$ & $0.910\pm0.005$ & $3.950\pm0.093$ & $12.756\pm0.181$ & $0.150\pm0.013$    & $1.306$ \\
8(d)    & Peripheral& $\pi^-$   & $0.073\pm0.004$ & $1.025\pm0.004$ & $0.398\pm0.008$ & $0.943\pm0.004$ & $2.653\pm0.091$ & $11.093\pm0.169$ & $10.627\pm0.888$   & $3.239$ \\
        &           & $K^-$     & $0.082\pm0.005$ & $1.033\pm0.005$ & $0.380\pm0.008$ & $0.891\pm0.005$ & $3.683\pm0.092$ & $12.553\pm0.170$ & $0.470\pm0.052$    & $3.570$ \\
        &           & $\bar{p}$ & $0.085\pm0.005$ & $1.009\pm0.005$ & $0.359\pm0.008$ & $0.910\pm0.005$ & $3.950\pm0.093$ & $13.018\pm0.181$ & $0.117\pm0.011$    & $0.647$ \\
\hline
9(c)    & Peripheral& $\pi^+$   & $0.089\pm0.004$ & $1.041\pm0.005$ & $0.446\pm0.010$ & $0.929\pm0.006$ & $2.403\pm0.075$ & $8.398\pm0.169$  & $14.318\pm0.567$   & $12.971$\\
Pb-Pb   &           & $K^+$     & $0.099\pm0.005$ & $1.065\pm0.005$ & $0.446\pm0.010$ & $0.926\pm0.006$ & $2.375\pm0.071$ & $7.468\pm0.153$  & $0.650\pm0.062$    & $1.544$ \\
$n_0=1$ &           & $p$       & $0.110\pm0.005$ & $1.030\pm0.005$ & $0.446\pm0.010$ & $0.894\pm0.007$ & $2.415\pm0.077$ & $8.005\pm0.161$  & $0.157\pm0.014$    & $2.881$ \\
9(d)    & Peripheral& $\pi^-$   & $0.089\pm0.004$ & $1.041\pm0.005$ & $0.446\pm0.010$ & $0.929\pm0.006$ & $2.403\pm0.075$ & $8.398\pm0.169$  & $14.318\pm0.567$   & $12.947$\\
        &           & $K^-$     & $0.099\pm0.005$ & $1.065\pm0.005$ & $0.446\pm0.010$ & $0.926\pm0.006$ & $2.375\pm0.071$ & $7.468\pm0.153$  & $0.650\pm0.062$    & $1.724$ \\
        &           & $\bar{p}$ & $0.110\pm0.005$ & $1.030\pm0.005$ & $0.446\pm0.010$ & $0.894\pm0.007$ & $2.415\pm0.077$ & $8.005\pm0.161$  & $0.157\pm0.014$    & $3.065$ \\
\hline
9(a)    &  Central  & $\pi^+$   & $0.099\pm0.005$ & $1.006\pm0.004$ & $0.435\pm0.006$ & $0.989\pm0.003$ & $2.775\pm0.085$ & $7.515\pm0.158$  & $1099.140\pm107.121$ & $2.897$ \\
Pb-Pb   &           & $K^+$     & $0.113\pm0.005$ & $1.002\pm0.001$ & $0.435\pm0.006$ & $0.984\pm0.003$ & $3.575\pm0.101$ & $7.735\pm0.115$  & $73.563\pm7.358$     & $3.623$ \\
$n_0=2$ &           & $p$       & $0.155\pm0.005$ & $1.002\pm0.001$ & $0.419\pm0.004$ & $0.996\pm0.002$ & $4.975\pm0.109$ & $8.225\pm0.128$  & $10.566\pm0.284$     & $15.778$\\
9(b)    &  Central  & $\pi^-$   & $0.099\pm0.005$ & $1.006\pm0.004$ & $0.435\pm0.006$ & $0.989\pm0.003$ & $2.775\pm0.085$ & $7.515\pm0.158$  & $1099.140\pm107.121$ & $2.955$ \\
        &           & $K^-$     & $0.113\pm0.005$ & $1.002\pm0.001$ & $0.435\pm0.006$ & $0.984\pm0.003$ & $3.575\pm0.101$ & $7.735\pm0.115$  & $73.563\pm7.358$     & $3.282$ \\
        &           & $\bar{p}$ & $0.155\pm0.005$ & $1.002\pm0.001$ & $0.419\pm0.004$ & $0.996\pm0.002$ & $4.975\pm0.109$ & $8.225\pm0.128$  & $9.983\pm0.278$      & $14.519$\\
9(c)    & Peripheral& $\pi^+$   & $0.079\pm0.004$ & $1.045\pm0.005$ & $0.405\pm0.008$ & $0.976\pm0.004$ & $3.003\pm0.095$ & $8.335\pm0.129$  & $14.692\pm0.760$     & $7.361$ \\
        &           & $K^+$     & $0.086\pm0.005$ & $1.053\pm0.005$ & $0.399\pm0.008$ & $0.928\pm0.004$ & $2.375\pm0.089$ & $7.475\pm0.121$  & $0.760\pm0.084$      & $0.975$ \\
        &           & $p$       & $0.102\pm0.005$ & $1.025\pm0.005$ & $0.385\pm0.007$ & $0.940\pm0.006$ & $2.675\pm0.092$ & $7.965\pm0.126$  & $0.177\pm0.014$      & $2.380$ \\
9(d)    & Peripheral& $\pi^-$   & $0.079\pm0.004$ & $1.045\pm0.005$ & $0.405\pm0.008$ & $0.976\pm0.004$ & $3.003\pm0.095$ & $8.335\pm0.129$  & $14.692\pm0.760$     & $7.488$ \\
        &           & $K^-$     & $0.086\pm0.005$ & $1.053\pm0.005$ & $0.399\pm0.008$ & $0.928\pm0.004$ & $2.375\pm0.089$ & $7.475\pm0.121$  & $0.760\pm0.084$      & $1.069$ \\
        &           & $\bar{p}$ & $0.102\pm0.005$ & $1.025\pm0.005$ & $0.385\pm0.007$ & $0.940\pm0.006$ & $2.675\pm0.092$ & $7.965\pm0.126$  & $0.171\pm0.014$      & $2.410$ \\
\hline
\end{tabular}
\end{center}}
\end{table*}

\begin{figure*}
\hskip-1.0cm \begin{center}
\includegraphics[width=16.0cm]{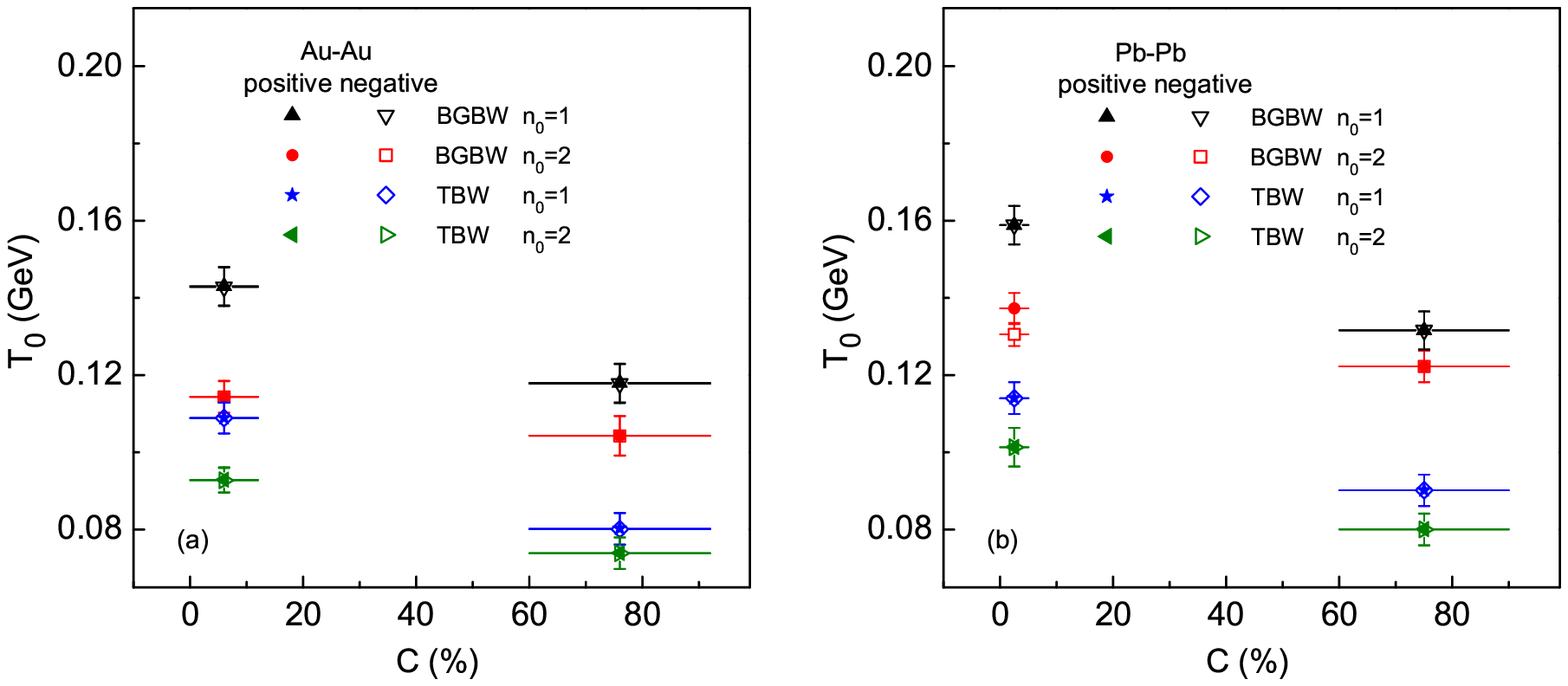}
\end{center}
\vskip0.5cm {\small Fig. 10. (Colour figure online) Comparisons of
$T_0$ obtained by the first two methods with $n_0=1$ and 2 for
different centralities. Panels (a) and (b) correspond to the
results for central (0--5\% and 0--12\%) and peripheral (80--92\%
and 60--80\%) Au-Au collisions at $\sqrt{s_{NN}}=200$ GeV and
central (0--5\%) and peripheral (80--90\% and 60--80\%) Pb-Pb
collisions at $\sqrt{s_{NN}}=2.76$ TeV respectively. The values of
$T_0$ are obtained by weighting different particles and the
results for central collisions obtained by the method i) with
$n_0=2$ and by the method ii) with $n_0=1$ are the same as Fig.
6.}
\end{figure*}

\begin{figure*}
\hskip-1.0cm \begin{center}
\includegraphics[width=16.0cm]{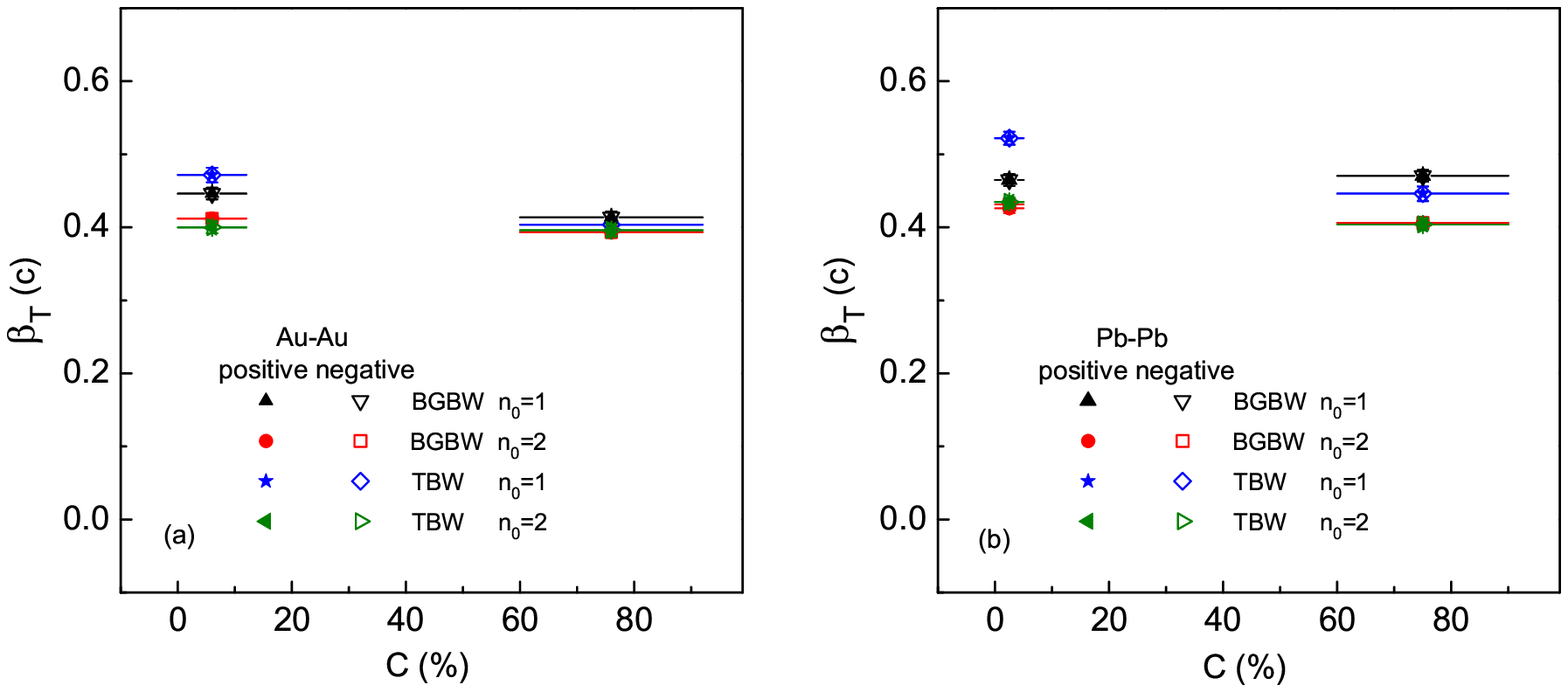}
\end{center}
\vskip0.5cm {\small Fig. 11. (Colour figure online) Same as Fig.
10, but showing the comparisons of $\beta_T$ obtained by the first
two methods for different centralities. The results for central
collisions obtained by the method i) with $n_0=2$ and by the
method ii) with $n_0=1$ are the same as Fig. 7.}
\end{figure*}

To give new comparisons for $T_0$ and $\beta_T$, the new results
obtained by the first two methods are shown in Figs. 10 and 11
respectively, where the results corresponding to the method i) for
central collisions with $n_0=2$ and to the method ii) for central
collisions with $n_0=1$ are the same as those in Figs. 6 and 7.
Combing Figs. 6, 7, 10, and 11, one can see that the four methods
show approximately the consistent results. These comparisons
enlighten us to use the first two methods in peripheral collisions
by a non-zero $\beta_T$. After the re-examination for $\beta_T$ in
peripheral collisions, we obtain a relative larger $T_0$ in
central collisions for the four methods. In particular, the
parameter $T_0$ at the LHC is slightly larger than or nearly equal
to that at the RHIC, not only for central collisions but also for
peripheral collisions. Except the method iii), other methods show
a slightly larger or nearly invariant $\beta_T$ in central
collisions comparing with peripheral collisions, and at the LHC
comparing with the RHIC, while the method iii) shows a nearly the
same $\beta_T$ in different centralities and at different
energies.

We would like to point out that, in the re-examination for
$\beta_T$ in the methods i) and ii), we have assumed both
$\beta_T$ in central and peripheral collisions to be non-zero.
While, in most cases [11, 14], both the conventional BGBW and TBW
models used non-zero $\beta_T$ in central collisions and zero (or
almost zero) $\beta_T$ in peripheral collisions. In the case of
using a non-zero or zero (or almost zero) $\beta_T$ in peripheral
collisions, we can obtain a relatively smaller or larger $T_0$,
comparing with central collisions. Indeed, the selection of
$\beta_T$ in peripheral collisions is an important issue in both
the BGBW and TBW models. In fact, $\beta_T$ is a sensitive
quantity which can affect $T_0$. The larger $\beta_T$ is selected,
the smaller $T_0$ is needed. The main correlation is between
$\beta_T$ and $T_0$, and the effect of $n_0$ is very small. In
Figs. 1 and 2, we have used a zero $\beta_T$ for peripheral
collisions and obtained a harmonious result on relative size of
$T_0$ with Ref. [28] in which $\beta_T$ ($0.35c$) for peripheral
collisions is nearly a half of that ($0.65c$) for central
collisions, and $n_0$ is also different from ours. While in Figs.
8 and 9, we have used a non-zero and slightly smaller $\beta_T$
for peripheral collisions and obtained a different result from
Ref. [28].

In order to make the conclusion more convincing, we can only fit
the low $p_T$ region of the particle spectra using the four
methods with the same $p_T$ cut to decrease the number of free
fitting parameters. When the $p_T$ cut increases from 2 to 3.5
GeV/$c$, $T_0$ (or $T$) increases or both $T_0$ (or $T$) and
$\beta_T$ increase slightly. The relative size of $T_0$
($\beta_T$) obtained above for central and peripheral collisions
is unchanged. In particular, $\beta_T$ is also a sensitive
quantity. For peripheral collisions, a zero or non-zero $\beta_T$
in the first two methods can give different results. In our
opinion, in central and peripheral collisions, it depends on
$\beta_T$ if we want to determine which $T_0$ is larger. We are
inclined to use a non-zero $\beta_T$ for peripheral collisions due
to small system which is similar to peripheral collisions in
number of participant nucleons also showing collective expansion
[40].

Comparing with peripheral collisions, the larger $T_0$ in central
collisions renders more deposition of collision energy and higher
excitation of interacting system due to more participant nucleons
taking part in the violent collisions. Comparing with the top RHIC
energy, the larger $T_0$ at the LHC energy also renders more
deposition of collision energy and higher excitation of
interacting system due to higher $\sqrt{s_{NN}}$ at the LHC. At
the same time, from the top RHIC to the LHC energies, a nearly
invariant $T_0$ reflects the limiting deposition of collision
energy. Comparing with peripheral collisions, the slight larger or
nearly the same $\beta_T$ in central collisions renders similar
expansion in both the centralities. At the same time, at the top
RHIC and the LHC energies, the two systems also show similar
expansion due to similar $\beta_T$.

It should be noted that, although Eq. (2) [14] does not implement
the azimuthal integral over the freeze-out surface which gives
rise to the modified Bessel functions in Eq. (1), it does not
affect the extractions of kinetic freeze-out parameters due to the
application of numerical integral. Although Eq. (3) [15, 16]
assumes a single, infinitesimally thin shell of fixed flow
velocity and also does not perform the integral over the
freeze-out surface, it can extract the mean trend of kinetic
freeze-out parameters. As for the alternative method [12, 17--20,
22--24], it assumes non-relativistic flow velocities in the
expressions used to extract the freeze-out parameters, which is
the case that $\beta_T$ is indeed not too large at the top RHIC
and LHC energies.
\\

{\section{Conclusions}}

We summarize here our main observations and conclusions.

(a) The $p_T$ spectra of $\pi^{\pm}$, $K^{\pm}$, $K_S^0$, $p$, and
$\bar p$ produced in central (0--5\% and 0--12\%) and peripheral
(80--92\% and 60--80\%) Au-Au collisions at $\sqrt{s_{NN}}=200$
GeV and in central (0--5\%) and peripheral (80--90\% and 60--80\%)
Pb-Pb collisions at $\sqrt{s_{NN}}=2.76$ TeV have been analyzed by
a few different superpositions in which the distributions related
to the extractions of $T_0$ and $\beta_T$ are used for the soft
component and the inverse power law is used for the hard
component. We have used five distributions, i) the Blast-Wave
model with Boltzmann-Gibbs statistics, ii) the Blast-Wave model
with Tsallis statistics, iii) the Tsallis distribution with flow
effect, iv)$_{\rm a}$ the Boltzmann distribution, and iv)$_{\rm
b}$ the Tsallis distribution, for the soft component. The first
three distributions are in fact three methods for the extractions
of $T_0$ and $\beta_T$. The last two distributions are used in the
fourth method, i.e. the alternative method.

(b) The experimental data measured by the PHENIX, STAR, and ALICE
Collaborations are fitted by the model results. Our calculations
show that the parameter $T_0$ obtained by the method i) with the
conventional treatment for central collisions is smaller than that
for peripheral collisions, which is inconsistent with the results
obtained by other model methods. In the conventional treatment,
the parameter $\beta_T$ in peripheral collisions is taken to be
nearly zero, which results in a larger $T_0$ than normal case. By
using the conventional treatment, both the methods i) and ii) show
a nearly zero $\beta_T$ in peripheral collisions according to
refs. [11, 14], while other methods show a considerable $\beta_T$
in both central and peripheral collisions.

(c) In central and peripheral collisions, we have to select a
suitable $\beta_T$ so that we can determine which $T_0$ is larger.
We are inclined to use a non-zero $\beta_T$ for peripheral
collisions due to small system also showing collective expansion.
We have given a re-examination for $\beta_T$ in peripheral
collisions in the methods i) and ii) in which $\beta_T$ is taken
to be $\sim(0.40\pm0.07)c$. By using a non-zero $\beta_T$, the
first two methods show approximately consistent results with other
methods, not only for $T_0$ but also for $\beta_T$, though the
method iii) gives a larger $\beta_T$. We have uniformly obtained a
larger $T_0$ in central collisions by the four methods. In
particular, the parameter $T_0$ at the LHC is larger than or equal
to that at the RHIC. Except the method iii), other methods show a
slightly larger or nearly invariant $\beta_T$ in central
collisions comparing with peripheral collisions, and at the LHC
comparing with the RHIC.

(d) The new results obtained by the widely used Blast-Wave model
with Boltzmann-Gibbs or Tsallis statistics are in agreement with
those obtained by the newly used alternative method which uses the
Boltzmann or Tsallis distribution. This consistency confirms the
validity of the alternative method. The result that the central
collisions have a larger $T_0$ renders more deposition of
collision energy and higher excitation of interacting system due
to more participant nucleons taking part in the violent
collisions. From the RHIC to LHC, the slightly increased or nearly
invariant $T_0$ renders the limiting or maximum deposition of
collisions energy. From central to peripheral collisions and from
the RHIC to LHC, the slightly increased or nearly invariant
$\beta_T$ renders the limiting or maximum blast of interacting
system.
\\

{\bf Conflict of Interests}

The authors declare that there is no conflict of interests
regarding the publication of this paper.
\\

{\bf Acknowledgments}

This work was supported by the National Natural Science Foundation
of China under Grant Nos. 11575103 and 11747319, the Shanxi
Provincial Natural Science Foundation under Grant No.
201701D121005, and the Fund for Shanxi ``1331 Project" Key
Subjects Construction.
\\

\end{multicols}
\end{document}